\pdfoutput=1
\documentclass[PAPER, atlasdraft=false, UKenglish, cernpreprint, texmf, orcidlogo]{atlasdoc}
\usepackage{atlaspackage}
\usepackage{atlasbiblatex}

\usepackage[hepprocess=true,jetetmiss=true]{atlasphysics}
\graphicspath{{logos/}{figures/}}

\usepackage{ANA-TOPQ-2018-19-PAPER-defs}

\AtlasTitle{Measurement of the top-quark mass using decays with a $J/\psi$ meson at $\sqrt{s}=$13 TeV with the ATLAS detector}
\AtlasAbstract{%
The top-quark mass is measured using top-quark decays producing an isolated
lepton and $J/\psi$ meson reconstructed in its $\mu^+\mu^-$ decay mode.
The data sample was recorded with the ATLAS detector in proton--proton
collisions at a centre-of-mass energy of $\sqrt{s}=13$~TeV during Run~2 of
the Large Hadron Collider, corresponding to an integrated luminosity of
140~$\textrm{fb}^{-1}$.
The measurement is based on the invariant mass $m(\ell \mu^+\mu^-)$ of the
system made of the isolated lepton $\ell$ from the $W$ boson decay and
the non-isolated $\mu^+\mu^-$ pair from a $J/\psi$ decay of a $b$-hadron,
exploiting its sensitivity to the top-quark mass.
An unbinned maximum-likelihood fit to the $m(\ell \mu^+\mu^-)$ distribution is
performed to extract the top-quark mass.
The top-quark mass is measured to be
$m_\textrm{top} = 172.17 \pm 0.80 \, \textrm{(stat)} \pm 0.81 \, \textrm{(syst)} \pm 1.07 \, \textrm{(recoil)}$~GeV,
with a total uncertainty of 1.56~GeV.
The third uncertainty arises from changing the dipole parton shower
gluon-recoil scheme used in top-quark decays.
}

\AtlasRefCode{TOPQ-2018-19}

\PreprintIdNumber{CERN-EP-2025-249}

\arXivId{2511.23091}

\HEPDataRecord{3087833}

\AtlasJournal{JHEP}
\AtlasJournalRef{JHEP 04 (2026) 099}
\AtlasDOI{10.1007/JHEP04(2026)099}
\AtlasCoverEgroupAnalysisTeam{atlas-topq-2018-19-analysis-team@cern.ch}

\hypersetup{pdftitle={ATLAS document},pdfauthor={The ATLAS Collaboration}}

\begin{document}

\maketitle

\tableofcontents

%
%

%

\section{Introduction}
The top quark is the heaviest fundamental particle observed so far and precise
knowledge of its mass (\mtop) is crucial to test the consistency of
the Standard Model (SM) of particle
physics~\cite{EWFit1,Stability1,Stability2,Stability3,Stability4}.
The top-quark mass is a renormalisation-scheme-dependent parameter
in perturbative quantum field theory.
The precise identification of the \mtop parameter in Monte Carlo (MC)
simulations within a field-theory mass scheme is the subject
of theoretical studies~\cite{Hoang:2018,Nason2019,Corcella2019,Azzi:2019,Hoang:2020,Hoang:2023}.
Since the discovery of the top quark at the Tevatron~\cite{CDF:1995,D0:1995},
the CDF and D0 collaborations have made multiple measurements of \mtop,
culminating in the 2016 combined result~\cite{CDFD0:2016}.
The two general-purpose experiments at the
Large Hadron Collider (LHC)~\cite{Evans:2008zzb},
ATLAS~\cite{PERF-2007-01} and CMS~\cite{CMS-CMS-00-001}, produced multiple
measurements of \mtop using data collected in proton–proton (\pp) collisions
at \rts~=~7~\TeV\ and 8~\TeV.
The combination of these measurements is
$\mtop = 172.52 \pm 0.14 \, \textrm{(stat)} \pm 0.30\, \textrm{(syst)}$~\GeV,
with a total uncertainty of $0.33$~\GeV~\cite{TOPQ-2019-13}.
It includes measurements with top-quark pair events that exploit
both the semileptonic and hadronic decays of the top quark, and a measurement
using events enriched in single-top-quark production via the electroweak
$t$-channel.
The uncertainty is dominated by the contribution of the \bjet and light-jet
energy scale uncertainties, respectively $0.18$ and $0.11$~\GeV.

Using the large data sample provided by the LHC during \RunTwo\ at
\rts~=~13~\TeV, the most recent ATLAS top-quark mass measurement was performed
in the lepton-plus-jets channel with a high transverse momentum top quark,
leading to
$\mtop = 172.95 \pm 0.27 \, \textrm{(stat)} \pm 0.46\, \textrm{(syst)}$~\GeV~\cite{TOPQ-2022-24}, while the most precise single measurement to date is
$171.77\pm 0.37$~\GeV~\cite{CMS-TOP-20-008} from the CMS Collaboration.
With a complementary approach, ATLAS performed a measurement using a partial
invariant mass reconstruction of the top-quark decay products where one of
the \bquarks hadronises into a \bhadron which decays semileptonically into
a muon~\cite{TOPQ-2017-17}, leading to
$\mtop = 174.41 \pm 0.39 \, \textrm{(stat)} \pm 0.66\, \textrm{(syst)} \pm 0.25\, \textrm{(recoil)}$~\GeV,
where the third uncertainty arises from changing the dipole parton shower
gluon-recoil scheme used in top-quark decays.

In this paper, a measurement of \mtop is presented using a partial
reconstruction of top quarks in final states that contain a \Jpsi
meson from a \bhadron decay.
Both top-quark--antiquark pair (\ttbar) and single-top-quark production are
considered to be signal in this study.
The decay mode of interest is $t\rightarrow(W\rightarrow\ell\nu) (b\rightarrow J/\psi+X \rightarrow\mu^+\mu^- + X)$,
the charge conjugation being implicit.
This channel has been suggested by the CMS Collaboration in
Ref.~\cite{CMSLetter} and refined in Ref.~\cite{Kharchilava} with a first
measurement leading to
$\mtop = 173.5 \pm 3.0 \, \textrm{(stat)} \pm 0.9\, \textrm{(syst)}$~\GeV~\cite{CMS-TOP-15-014}.
The key motivation for using \Jpsi events lies in their reduced dependence
on jet reconstruction.
An unbinned maximum-likelihood fit to an observable sensitive to
\mtop\ is performed.
This method exploits the invariant mass
\mlmumu, built from the isolated lepton $\ell$ (with $\ell=e,\mu$) from the
\Wboson-boson decay and the $\mu^+\mu^-$ pair from the \Jpsi candidate.
As this observable consists entirely of leptons, it benefits
from improved momentum resolution and a reduction of sensitivity to the
systematic uncertainties related to jet energy calibration.
However, this method presents some challenges: a small branching ratio of the
signal process and sensitivity to the signal process modelling,
especially parton shower and hadronisation, \bquark fragmentation,
and radiation.
Methods with different types of systematic uncertainties are
important when combining measurements, and for testing the consistency of the
theoretical interpretation of the top-quark mass.
%


%

%
\newcommand{\AtlasCoordFootnote}{%
ATLAS uses a right-handed coordinate system with its origin at the nominal interaction point (IP)
in the centre of the detector and the \(z\)-axis along the beam pipe.
The \(x\)-axis points from the IP to the centre of the LHC ring,
and the \(y\)-axis points upwards.
Polar coordinates \((r,\phi)\) are used in the transverse plane,
\(\phi\) being the azimuthal angle around the \(z\)-axis.
The pseudorapidity is defined in terms of the polar angle \(\theta\) as \(\eta = -\ln \tan(\theta/2)\) and is equal to the rapidity
$ y = \frac{1}{2} \ln \left( \frac{E + p_z}{E - p_z} \right) $ in the relativistic limit.
Angular distance is measured in units of \(\Delta R \equiv \sqrt{(\Delta y)^{2} + (\Delta\phi)^{2}}\).}
\section{ATLAS detector}
The ATLAS detector~\cite{PERF-2007-01} at the LHC covers nearly the entire solid angle around the collision point.\footnote{\AtlasCoordFootnote}
It consists of an inner tracking detector surrounded by a thin superconducting solenoid, electromagnetic and hadronic calorimeters,
and a muon spectrometer incorporating three large superconducting air-core toroidal magnets.
The inner-detector system (ID) is immersed in a \qty{2}{\tesla} axial magnetic field
and provides charged-particle tracking in the range \(|\eta| < 2.5\).
The high-granularity silicon pixel detector covers the vertex region and typically provides four measurements per track,
the first hit generally being in the insertable B-layer (IBL) installed before Run~2~\cite{ATLAS-TDR-19,PIX-2018-001}.
It is followed by the SemiConductor Tracker (SCT), which usually provides eight measurements per track.
These silicon detectors are complemented by the transition radiation tracker (TRT),
which enables radially extended track reconstruction up to \(|\eta| = 2.0\).
The TRT also provides electron identification information
based on the fraction of hits (typically 30 in total) above a higher energy-deposit threshold corresponding to transition radiation.
The calorimeter system covers the pseudorapidity range \(|\eta| < 4.9\).
Within the region \(|\eta|< 3.2\), electromagnetic calorimetry is provided by barrel and
endcap high-granularity lead/liquid-argon (LAr) calorimeters,
with an additional thin LAr presampler covering \(|\eta| < 1.8\)
to correct for energy loss in material upstream of the calorimeters.
Hadronic calorimetry is provided by the steel/scintillator-tile calorimeter,
segmented into three barrel structures within \(|\eta| < 1.7\), and two copper/LAr hadronic endcap calorimeters.
The solid angle coverage is completed with forward copper/LAr and tungsten/LAr calorimeter modules
optimised for electromagnetic and hadronic energy measurements, respectively.
The muon spectrometer (MS) comprises separate trigger and
high-precision tracking chambers measuring the deflection of muons in a magnetic field generated by the superconducting air-core toroidal magnets.
The field integral of the toroids ranges between \num{2.0} and \qty{6.0}{\tesla\metre}
across most of the detector.
Three layers of precision chambers, each consisting of layers of monitored drift tubes, cover the region \(|\eta| < 2.7\),
complemented by cathode-strip chambers in the forward region, where the background is highest.
The muon trigger system covers the range \(|\eta| < 2.4\) with resistive-plate chambers in the barrel, and thin-gap chambers in the endcap regions.
The luminosity is measured mainly by the LUCID--2~\cite{LUCID2} detector that records Cherenkov light produced in the quartz windows of photomultipliers located close to the beampipe.
Events were selected by the first-level trigger system implemented in custom hardware,
followed by selections made by algorithms implemented in software in the high-level trigger~\cite{TRIG-2016-01}.
The first-level trigger accepted events from the \qty{40}{\MHz} bunch crossings at a rate close to \qty{100}{\kHz},
which the high-level trigger further reduced in order to record complete events to disk at about \qty{1.25}{\kHz}.
A software suite~\cite{SOFT-2022-02} is used in data simulation, in the reconstruction
and analysis of real and simulated data, in detector operations, and in the trigger and data acquisition
systems of the experiment.


%

\section{Data and simulated event samples}
\label{sec:datasimu}
The data were recorded with the ATLAS detector during \RunTwo of
the LHC, which took place from 2015 to 2018,
in \pp collisions at a centre-of-mass energy of
$\rts=13$~\TeV, corresponding to an integrated luminosity of 140~\ifb.
Only events recorded under stable beam conditions with all detector subsystems
operational~\cite{DAPR-2018-01} are used.
MC simulated event samples are used in the analysis to model
the \ttbar and single-top-quark events and most of the other processes,
except non-prompt-lepton and fake-lepton
backgrounds, which are estimated by using data-driven methods.
The MC samples were processed either through a full simulation of the detector
response~\cite{SOFT-2010-01} with the \GEANTFOUR\
toolkit~\cite{Agostinelli:2002hh}, or using the Atlfast2 fast
simulation with a parameterised response of the calorimeter.
All simulated samples were overlaid with additional \pp interactions (pile-up),
generated with \PYTHIA[8.186]~\cite{Sjostrand:2007gs} using the
\NNPDF[2.3lo]~\cite{Ball:2012cx} set of parton distribution functions (PDF)
and the A3 set of tuned parameters~\cite{ATL-PHYS-PUB-2016-017} (tune).
The average number of interactions per bunch crossing is reweighted to match
that in data.
Events were processed using the same reconstruction algorithms as applied
to data.
An \mtop value of 172.5~\GeV\ was used in all simulated samples,
unless stated otherwise.

The production of \ttbar\ events was simulated using the
\POWHEGBOX~v2~\cite{Nason:2004rx,Frixione:2007vw,Alioli:2010xd,Frixione:2007nw}
generator, which provides matrix elements (ME) at next-to-leading order (\NLO)
in quantum chromodynamics (QCD), and the \NNPDF[3.0nnlo]~\cite{Ball:2014uwa}
set.
The \hdamp parameter, which effectively regulates the high transverse
momentum (\pt) radiation against which the \ttbar system recoils,
was set to 1.5\ensuremath{m_{\textrm{top}}}~\cite{ATL-PHYS-PUB-2016-020}.
The functional form of the renormalisation (\muR) and factorisation (\muF)
scales was set to
$\muR=\muF=\sqrt{(\mtop^2+p^{2}_{\textrm{T},t})}$, where the top-quark transverse
momentum $p_{\textrm{T},t}$ was evaluated before
radiation~\cite{ATL-PHYS-PUB-2016-004}.
Parton showering (PS) and hadronisation was modelled with
the \PYTHIA[8.230]~\cite{Sjostrand:2014zea} generator using the
A14 tune~\cite{ATL-PHYS-PUB-2014-021} and the \NNPDF[2.3lo] set.
ME corrections that approximate \NLO\ QCD were enabled in \PYTHIA
for all emissions, compensating for the leading-order (\LO) precision used in
\POWHEGhvq to simulate the top-quark decay.
The \pthard\ parameter in \PYTHIA that affects the matching of the PS
to the ME calculation was set to zero.
The recoil target for secondary gluon emissions from the \bquark in the
$t\rightarrow Wb$ vertex, was assigned to the
\bquark.\footnote{This corresponds to the setting
\textit{recoil-to-colour=ON} in the TimeShower in \PYTHIA.}

The A14 tune of \PYTHIA[8] is based on the \Monash tune~\cite{Skands:2014pea}
of PS and multiple parton interactions (MPI) parameters
that leaves the hadronisation parameters at their default values and uses
the Lund--Bowler fragmentation model~\cite{LundModel2}:
\begin{equation*}
f(z) = \frac{1}{z^{1+br_bm_{b}^2}} (1-z)^a \exp(-bm_{\textrm{T}}^2/z) \, ,
\end{equation*}
where $a$, $b$ and \rb are the function parameters, $m_b$ is the \bquark
mass, $m_{\textrm{T}}=\sqrt{m_B^2+p_{\textrm{T},B}^2}$ is the \bhadron
transverse mass ($m_B$ and $p_{\textrm{T},B}$ being the \bhadron mass
and transverse momentum, respectively) and $z$ is the fraction
of the longitudinal energy of the \bhadron relative to the \bquark,
in the light cone reference frame.
The values of $a$ and $b$ were fit to data sensitive to
light-quark fragmentation such as charged-particle
multiplicities, event shapes and scaled momentum distributions.
They are assumed to be universal for light- and heavy-quarks, while the
\rb parameter is specific to \bquark fragmentation.
The A14 tune sets the value of the strong coupling constant in the final-state
shower (\alphasFSR) to \num{0.127}, whereas the value
of \num{0.1365} is used in Monash.
However, both Monash and A14 set $\rb=0.855$.
In Ref.~\cite{TOPQ-2017-17}, a more appropriate \rb value was fitted
for a value of $\ensuremath{\alpha_{\text{s}}^{\text{FSR}}}=0.127$.
The fit used the A14 tune with $e^+e^-$ collision data from
LEP/SLD~\cite{AlephbFrag,DelphibFrag,OpalbFrag,SLDbFrag}.
The distribution
$\xB=2p_B \cdot p_Z / m_{Z}^{2}$ from semileptonically decaying
\bhadrons in $e^+e^-\rightarrow Z\rightarrow b\bar{b}$ events was used, where
$p_B$ and $p_Z$ are the four-momenta of the \bhadron and the \Zboson boson
respectively.
A value of $\rb=1.05\pm 0.02$ was found to be optimal and the central value
defines the A14--\rb tune used by default in this analysis.
This tune provides good agreement with collision data in a measurement of
\bquark fragmentation in top-quark decays~\cite{TOPQ-2017-19}.

The single-top-quark events are split into three processes:
$s$-channel, $t$-channel and \tW associated production.
These samples were simulated using the \POWHEGBOX~v2 generator
at \NLO\ in QCD using the four-flavour (five-flavour) scheme
for the $t$-channel ($s$-channel and \tW) with the
\NNPDF[3.0nnlo] set.
For the $t$-and $s$-channels, the functional form of \muR and \muF was set
to $\sqrt{(m_{b}^2+p_{\textrm{T},b}^2)}$, where $p_{\textrm{T},b}$ is the \bquark
transverse momentum.
The interference between the \ttbar\ and the \tW\ final states was handled
using the diagram removal scheme~\cite{Frixione:2005vw,Frederix:2012dh}.
Events were further processed with \PYTHIA[8.230]
using the A14--\rb tune and the \NNPDF[2.3lo] set.

The production fractions of weakly decaying \bhadrons observed in
\POWPY MC simulation were rescaled to those from the Heavy
Flavour Averaging Group (HFLAV)~\cite{HFAV2018} as reported by the
Particle Data Group (PDG)~\cite{Zyla:2020zbs} with the method
described in Ref.~\cite{ATL-PHYS-PUB-2022-035}.
The production fractions and corresponding scale factors
for \POWPY simulations are shown in Table~\ref{tab:sec:datasimu:hfprod}.
These scale factors refer only to the first weakly decaying hadron produced
in the hadronisation process of \bquarks.
The scale factors were applied to each of these hadrons present in a MC
simulated event, with the overall event weight given by the product
of these scale factors.
This procedure assumes that the production fractions of heavy-flavour hadrons
can be regarded as universal in the kinematic phase space relevant to this
analysis, within the uncertainties accounted for here, as supported by recent
results~\cite{LHCbbFrag1,LHCbbFrag2,LHCbbFrag3,LHCbbFrag4,LHCbbFrag5,LHCbbFrag6,ALICEbFrag} that are consistent with universality except for \pp collisions
with high charged-particle multiplicity at transverse momenta below 6~\GeV.
The assumption is that production fractions are the same in all experiments
and that they sum up to~one.
\begin{table}[!htb]
\centering
\caption{The production fraction values for \bhadrons in the
PDG and in \POWPY[8.2].
The relative scale factors applied to \POWPY[8.2] are also shown.
The values in the PDG column are derived from Ref.~\cite{Zyla:2020zbs}.
}
\label{tab:sec:datasimu:hfprod}
\begin{tabular}{cccc}
\toprule
Hadron         & PDG     & \POWPY[8.2] & Scale factor \\
\midrule
\Bzero         & $0.408\pm 0.006$  & 0.429  & 0.951 \\
\Bplus         & $0.408\pm 0.006$  & 0.429  & 0.951 \\
\Bs            & $0.100\pm 0.008$  & 0.094  & 1.064 \\
\bbaryon       & $0.084\pm 0.011$  & 0.047  & 1.787 \\
\Bc            & $0.0030\pm 0.0005$ & 0.0002 & 14.941 \\
\bottomrule
\end{tabular}
\end{table}

The \EVTGEN~\cite{Lange:2001uf} program was used to simulate bottom and charm
hadron mixing and decays.
The branching ratios ($\mathcal{BR})$ of the decay of \bhadrons into \jpsi
were also rescaled to the latest values from the PDG\@.
The total branching ratio of the decay of \bhadrons into \jpsi is
$\mathcal{BR}(b\rightarrow\jpsi+X) = (1.16\pm 0.1)\times 10^{-2}$.
The \jpsi can be produced directly with a
$\mathcal{BR}(b\rightarrow\jpsi(\textrm{direct})+X) = (7.8\pm 0.4)\times 10^{-3}$.
Indirect production appears through excited mesonic or baryonic states that
first deexcite with the emission of a photon and then decay into a \Jpsi.
The main processes are through the production of a $\psi(2S)$
and a $\chi_{c1}(1P)$ with branching fractions of
$(1.76\pm0.19)\times 10^{-3}$ and $(4.8\pm1.5)\times 10^{-3}$ respectively.
The \jpsi can decay into a $\mu^+\mu^-$ pair with a
$\mathcal{BR}(\jpsi\rightarrow\mu^+\mu^-)=(5.96\pm 0.03)\times 10^{-2}$.
Thus, the branching ratio of the decay of \bhadrons into \Jmm is
$\mathcal{BR}(b\rightarrow\jpsi(\rightarrow\mu^+\mu^-)+X) \sim 6.9\times 10^{-4}$.

The \ttbar and single-top-quark event samples were generated for different
assumed values of the top-quark mass, \mtopgen,
namely 169, 171, 172, 172.25, 172.5, 172.75, 173, 174 and 176~\GeV.

Alternative \ttbar and single-top-quark samples were simulated to estimate
the systematic uncertainties in the signal processes production modelling.
Several alternative samples varied just one of the parameters of the MC
generators, allowing individual systematic effects to be investigated
separately.
The ambiguities in matching the ME calculation and the PS are
tested with a sample where the \pthard\ parameter was set
to~1~\cite{ATL-PHYS-PUB-2023-029}.
The \ttbar production threshold may be sensitive to the modelling of
off-shell effects and top-quark decay.
To estimate the uncertainty in the modelling of these effects,
an alternative sample was used, where the top-quark decay was simulated with
the \MADSPIN generator in the \MADGRAPH framework~\cite{Artoisenet:2012st}
interfaced to \POWHEGBOX~v2.
The effect of using a different PS and hadronisation model is evaluated using
a \ttbar sample produced with \POWHEGBOX~v2 interfaced to
\HERWIG[7.1.3]~\cite{Bahr:2008pv,Bellm:2015jjp} using the H7-UE set of tuned
parameters and the \MMHT[lo]~\cite{Harland-Lang:2014zoa} set of PDFs.
For single-top-quark events the $t$- and \tW-channels samples were
simulated using \HERWIG[7.2.1].
To evaluate the uncertainty in the modelling of the \bquark fragmentation,
two additional samples were produced with the value of the \rb parameter
entering the Lund--Bowler parameterisation in the fragmentation function
varied by its uncertainty of $\pm 0.02$.
Another set of \ttbar and single-top-quark samples was produced with the
\hdamp parameter doubled to $3\mtop$.
To probe uncertainties in final-state QCD radiation (FSR), two sets of
\ttbar and single-top-quark samples were generated with the scale parameters
explicitly\footnote{With explicit scale variations, dedicated alternative MC
samples were generated with the $\mu_\textrm{r, f}$ scales modified in the
parton shower settings, as opposed to the automated parton shower variations
discussed in Ref.~\cite{Mrenna:2016sih}, where weights are applied
to the MC sample to obtain the systematic shifts.
Explicitly changing $\mu_\textrm{r, f}$ by a factor of $\sqrt{2}$ corresponds
approximately to an automated variation of a factor of two, thanks to the
implementation of the \NLO\ compensation terms in the latter case.}
varied up and down by factors of $\sqrt{2}$~\cite{Mrenna:2016sih}.
For these alternative settings, the appropriate \rb values were determined
following the same fit procedure described above in order to still correctly
model the \xB distribution for LEP/SLD data~\cite{TOPQ-2017-17}.
Uncertainties in the modelling of the underlying event in \ttbar events
are evaluated with two samples using the same algorithms for ME and PS as in
the nominal one, but using the Var1 settings~\cite{ATL-PHYS-PUB-2014-021}
from the A14 set of tuned parameters.
The parameters varied for Var1 up and down were the \alphas value in the
MPI model and the colour reconnection (CR) `range' parameter in \PYTHIA,
which steers the number of possible reconnections.
Three \ttbar samples were generated with alternative MPI and CR models and
tunes, denoted by CR0, CR1 and CR2~\cite{ATL-PHYS-PUB-2017-008}.
CR0 denotes a tune that uses the 'MPI-based CR model' and is essentially the
same as that used in the nominal sample, with the only difference being a
retuning of the colour-reconnection probability.
CR1 denotes a tune that uses a 'QCD-based model' while CR2 denotes a tune that
uses a 'gluon-move model'.
The ambiguity in the choice of the recoil particle for the secondary gluon
emission from a \bquark produced in $t\rightarrow Wb$ is addressed by
generating a sample where the top quark takes part in the recoil
(\PYTHIA parameter \textit{recoil-to-top}).

Unless stated otherwise, all cross-section values are stated for
\pp collisions at $\rts=13$~\TeV\ and for $\mtop= 172.5$~\GeV.
All \ttbar samples were normalised to the \mtop-dependent cross-section
prediction obtained from \TOPpp[2.0]~\cite{Czakon:2011xx} at
next-to-next-to-leading order (\NNLO) in QCD, including a resummation of
next-to-next-to-leading logarithmic (NNLL) soft gluon
terms~\cite{Cacciari:2011hy,Baernreuther:2012ws,Czakon:2012pz,Czakon:2012zr,Czakon:2013goa}.
The \ttbar cross-section was calculated to be
$\sigma(\ttbar)_\textrm{NNLO+NNLL} = 832\pm51$~pb.
The predicted cross-sections of single-top-quark and single-top-antiquark
production in the $t$-channel are
$\sigma(t,t\textrm{-chan})_\textrm{NNLO}=134.2^{+2.6}_{-1.7}$~pb
and
$\sigma(\bar{t},t\textrm{-chan})_\textrm{NNLO}=80.0^{+1.8}_{-1.4}$~pb,
respectively.
The cross-sections were calculated with the MCFM
program~\cite{Campbell:2020fhf} at \NNLO in QCD.
The quoted uncertainties for the $t$-channel include the uncertainties due to
the choice of the \muR and \muF scales, the uncertainty in the PDFs and in
the value of the strong coupling constant \alphas.
The scale uncertainty is determined by varying \muR and \muF independently
up and down by a factor of two, whilst never allowing them to differ by a
factor greater than two from each other -- the so-called independent restricted
scale variations.
The combined PDF and \alphas uncertainties are determined at the 68\%
confidence level according to the Hessian representation of the PDF4LHC21
set~\cite{Ball:2022oua}.
The predicted cross-section of \tW production is
$\sigma(t,\tW\textrm{-chan})_\textrm{NNLO+NNLL}=79.3^{+2.9}_{-2.8}$~pb and was
computed at \NNLO in QCD with the addition of third-order corrections of
soft-gluon emissions by resuming NNLL terms~\cite{Kidonakis:2021vob} using
the PDF4LHC21 set.
The quoted uncertainty for the \tW channel includes the uncertainty due to the
choice of \muR and \muF and the uncertainty in the PDFs.
The uncertainty in the scale choice is determined by varying the scales
simultaneously up and down by a factor of two.
The PDF uncertainties are based on the Hessian method and include
the uncertainty in \alphas.
The predicted cross-section of $s$-channel production is $\sigma(t,s\textrm{-chan})_\textrm{NLO}=11.07^{+0.19}_{-0.18}$~pb and was computed at \NNLO in
QCD~\cite{Liu:2018gxa}.
The PDF4LHC21 set was used.
The quoted uncertainty for the $s$-channel includes the uncertainty due to the
choice of $\mu_{r}$ and $\mu_{f}$ and the uncertainty in the PDFs.
The uncertainty in the scale choice was determined by varying the scales
simultaneously up and down by a factor of two.
The PDF uncertainties are based on the replicas method.

The production of \ttbar in association with one or two vector bosons (\ttV)
or with a Higgs boson (\ttH) was simulated using \MGNLO~(v2.3.3)~\cite{Alwall:2014hca}.
The generator provides ME calculation at \NLO\ in \alphas, with the
\NNPDF[3.0nnlo] set.
It was interfaced to \PYTHIA[8.210] and used the A14 set of tuned parameters
and the \NNPDF[2.3lo] set~\cite{ATL-PHYS-PUB-2016-005}.
Other background processes include production
of pairs of vector bosons (diboson, $WW$, $ZZ$, $WZ$) or \Wboson and \Zboson
bosons in association with jets, as well as specifically \Wboson bosons
in association with \Jpsi.
\SHERPA~v2.2.1 and v2.2.2 were used for diboson processes with two
and three leptons in the final state, respectively.
\SHERPA~v2.2.11~\cite{Bothmann:2019yzt} was used to model
\Wjets and Drell--Yan $\Zboson/\gamma^{\ast}$+jets production.
For these processes, \SHERPA calculates the ME at \NLO for up to two
partons and at leading order (\LO) for up to five partons using the
\OPENLOOPS~\cite{Cascioli:2011va} and \COMIX~\cite{Gleisberg:2008fv}
ME event generators.
They were matched with the \SHERPA PS~\cite{Schumann:2007mg}
using the MEPS\@NLO prescription.
It employs a dedicated set of tuned parameters developed by the \SHERPA
authors based on the \NNPDF[3.0nnlo] set.
The \MGNLO[2.6.5] program was used to generate alternative
$W$/\Zjets samples with up to three additional partons in the final state
at \NLO accuracy.
The showering and subsequent hadronisation were performed using \PYTHIA[8.245]
with the A14 tune and the \NNPDF[2.3lo] set with $\alphas = 0.13$.
The different jet multiplicities were merged using the \FXFX \NLO matrix-element
and parton-shower merging prescription~\cite{Frederix:2012ps}.
\PYTHIA[8.245] was used to model the PS, hadronisation
and underlying event.
These samples are referred to as \MGFXFX $W$/$Z$+jets samples in the following.
Two simulated samples of associated prompt \WJpsi production are
used for this analysis.
\MADONIA, a heavy quarkonia package~\cite{MadOnium} that runs on \MADGRAPH[4],
was used to generate colour-octet samples of single parton scattering (SPS)
associated production.
Les Houches Event Files (LHEF)~\cite{Alwall:2006yp} were generated using
$\textrm{MG\_ME\_V4.5.2}$ with the CTEQ6L1 set of PDFs~\cite{Boos:2001cv}.
The hard scatter LHEF from \MADGRAPH were then
interfaced to \PYTHIA[8.186] (using the AU2 tune~\cite{ATL-PHYS-PUB-2012-003})
for hadronisation and showering.
It is possible for the \Wboson boson and \Jpsi to originate from two different
parton interactions in the same \pp collision, in a double parton scattering
process (DPS).
Such a DPS sample was generated, hadronised, and
showered with \PYTHIA[8.186] using the AU2 tune and the CTEQ6L1 set.
Production of \ZJpsi is expected to be negligible.

The non-prompt- and fake-lepton background
corresponds to events originating from processes that do not involve isolated
leptons produced by \Wboson, \Zboson or Higgs boson decays.
It is estimated by using the data-driven ‘matrix-method’~\cite{EGAM-2019-01},
where data events passing the full selection with looser identification and
isolation requirements on leptons are used along with
the measured probability of loose leptons to satisfy the nominal
selection requirements.
%


%

\section{Object definition}
The primary vertex is defined as the vertex with the largest
$\sum\pTsq$ of the associated tracks among all the vertices
with at least two tracks with $\pt>0.5$~\GeV consistent with the
beam-collision region in
the $x$–$y$ plane~\cite{ATL-PHYS-PUB-2015-026}.

Electron candidates are reconstructed by matching energy deposits in the
electromagnetic calorimeter with a corresponding track in the inner
tracking detector~\cite{EGAM-2018-01}.
Electrons are required to satisfy the \textsc{TightLH} identification
criteria~\cite{EGAM-2018-01} and to lie within $\aeta<2.47$,
excluding candidates within the calorimeter transition region $1.37<\aeta<1.52$.
Muon candidates are reconstructed by combining tracks from the inner detector
with matching tracks reconstructed in the muon spectrometer~\cite{MUON-2018-03}.
They are required to have $\aeta<2.5$ and to satisfy the \textsc{Medium}
requirements~\cite{MUON-2018-03}.
Both lepton flavours are required to have a transverse momentum \pt greater
than 25~\GeV\ (27~\GeV\ for 2016--2018 data periods)
to ensure high efficiency of the single-lepton triggers.
Their tracks are matched to the primary vertex, by requiring that the
transverse impact parameter divided by its estimated uncertainty
be less than five (three) for electron (muon) candidates and
the longitudinal impact parameter is required to satisfy $|z_{0}(\ell)|<0.5$~mm.

Charged leptons originating from the \Wboson~boson decay have to satisfy
isolation requirements.
Isolated electron candidates are required to satisfy the \textsc{Tight}
isolation requirement~\cite{EGAM-2018-01} that removes fake and non-prompt
electrons from heavy-flavour decays.
Isolated muon candidates are required to satisfy the
\textsc{PflowTight\_FixedRad} isolation selection~\cite{MUON-2018-03}
to reduce background from heavy-flavour decays inside jets.

Soft-muons (referred to as soft-$\mu$) are a special case of muons that are
used to select \Jpsi in the final state.
They are reconstructed similarly to isolated muons but selected with a low
transverse momentum threshold $\pt>3$~\GeV.
To remove muons not originating from \pp collision events,
the longitudinal impact parameter is required to satisfy $|z_{0}(\mu)|<10$~mm.
The soft-$\mu$ are required to satisfy the \textsc{LowPt} working point
identification criterion~\cite{ATL-PHYS-PUB-2020-002} optimised to provide
good muon reconstruction efficiency down to 3~\GeV,
while controlling the fake-$\mu$ rate.
No isolation requirement is applied.
Soft-$\mu$ are associated with jets whenever the angular separation
$\dR$ is smaller than 0.4

Jet candidates are reconstructed from particle-flow
objects~\cite{PERF-2015-09} using the anti-$k_{t}$ jet
algorithm~\cite{Cacciari200657,Cacciari:2008gp} implemented in the FastJet
package~\cite{Fastjet} with a radius parameter $R=0.4$.
Jets are accepted if they have $\pt>25$~\GeV\ and $\aeta<2.5$.
To reduce the contribution from pile-up jets, jets with $\pt<60$~\GeV and
$\aeta<2.4$ are required to satisfy a criterion of the jet--vertex
tagger (\JVT) output value, $\JVT>0.59$.
The \JVT is a multivariate quantity that combines information from several
track-based variables~\cite{PERF-2014-03}.
Further corrections, derived from a calibration based on MC
simulations and data~\cite{JETM-2018-05}, are applied to the jet energy leading
to corrections of the jet four-momentum.
Jets containing \bhadrons are identified (‘$b$-tagged‘) using the
multivariate algorithm DL1r~\cite{FTAG-2019-07}, which
combines information about the impact parameters of displaced tracks
and the topological properties of secondary and tertiary decay vertices
reconstructed within the jet.
The algorithm is trained on simulated \ttbar events to discriminate \bjets
from a background consisting of light-flavour jets and \cjets.
The working point used has an efficiency of 77\% and a corresponding rejection
factor of 5 and 200 for $c$-quark jets and light-quark originated jets.

The missing transverse momentum $\vec{p}_\text{T}^\text{miss}$
(with magnitude \met), is defined as the negative vector sum of the
$\vec{p}_\text{T}$ of all selected and calibrated objects in the event,
including a term to account for the momenta of soft particles that are not
associated with any of the selected objects~\cite{JETM-2020-03}.
This soft term is calculated from ID tracks matched to the primary vertex,
which makes it more resilient to contamination from pile-up interactions.

To avoid double-counting, overlapping physics objects are rejected
in the following order: electrons sharing a track with a muon; the closest
jet within $\dR=0.2$ of an electron; electrons within $\dR=0.4$ of a jet;
jets within $\dR=0.4$ of a muon if they have at most two associated
tracks; muons within $\dR=0.4$ of a jet.
Soft-$\mu$ candidates matching ($|\Delta\eta|<0.001$ and
$|\Delta\phi|<0.001$) isolated muon candidates used to preselect
the ttbar events are rejected.


%

\section{Event selection}
\label{sec:sel}
The event selection targets top-quark events
in the final state $\ell \nu b qq' \bar{b}$, where $\ell=e$ or $\mu$ and
$qq'$ are the quarks from the \Wboson decay that is producing jets.
At least one $b$-initiated jet is required, along with a
\Jmm decay originating from a \bhadron.

Events are required to have at least one primary vertex and to have satisfied
a single-electron or single-muon trigger.
Multiple triggers were used to increase the selection efficiency.
The lowest-threshold triggers utilised isolation requirements to reduce the
trigger rate.
These had \pt thresholds of 20~\GeV\ for muons and 24~\GeV\ for electrons
in 2015 data, and 26~\GeV for both lepton types in 2016, 2017 and 2018
data~\cite{TRIG-2016-01,TRIG-2018-01,TRIG-2018-05}.
Events are selected by requiring exactly one isolated lepton candidate
matched to the trigger lepton and at least two jets with at least one
of them being $b$-tagged.
This loose requirement on jet multiplicity retains signal events
even at the expense of a larger background level.
Events containing additional isolated leptons with $\pt>25$~\GeV\ are rejected.
This analysis does not require the presence of missing transverse momentum.
About 35 million events satisfy this selection.

\Jpsi\ candidates are reconstructed using all pairs of oppositely charged
soft-$\mu$ that are constrained to originate from a common vertex using
ID track parameters.
The properties of the \Jpsi, such as its invariant mass, transverse
momentum, and pseudorapidity are determined from the result of the vertex fit.
The \Jpsi\ candidates consist of genuine prompt and non-prompt decays,
background decays produced by various combination of fake and real muons, and
genuine muon pairs producing an invariant mass in the continuum under
the peak.
The various components can be separated using the pseudo-proper lifetime
\tauzero of the \Jpsi\ candidates defined by
$\tauzero = \Lxy\times\mjpsiPDG/\ptjpsi$, with
$\Lxy = \vec{L}\cdot\vec{p_\textrm{T}}^\textrm{\Jpsi}/\ptjpsi$,
where $\vec{L}$ is the vector of the distance between the primary vertex and
the extrapolated common vertex of the two muon candidates in the
transverse plane, \mjpsiPDG is the mass of the \Jpsi\ from
the PDG, and $\vec{p_\textrm{T}}^\textrm{\Jpsi}$ is the reconstructed
transverse momentum vector of the \Jpsi\ candidate.
The mass variable \mjpsiPDG is used instead of the reconstructed value
in order to have uncorrelated mass and lifetime information.
Prompt \Jpsi\ decays have a pseudo-proper lifetime consistent with zero within
resolution.

Further \Jpsi selection criteria are applied:
\begin{itemize}
\item[-] The invariant mass of the \Jpsi candidate lies in the final
range of 2.9–3.3~\GeV.
A total of \num{19576} candidates remain in the data.
\item[-] The quality of the fit of the two soft-$\mu$ to a common vertex has
a $\chi^2<10$, for the one degree of freedom.
A total of \num{18133} candidates remain.%
\item[-] The \Jpsi candidate has a transverse momentum $\pt>8$~\GeV
and rapidity $|y|<2.1$.
A total of \num{17280} candidates remain.%
\item[-] The angular distance between the two soft-$\mu$ is $\dRmm<0.6$.
Both are inside the same jet and their angular distance to this jet
is $\dRsoftmjet<0.4$.
A total of \num{16821} candidates remain.%
\item[-] The proper lifetime of the candidates is $\tauzero>0.1$~ps.
A total of \num{15350} candidates remain.%
\end{itemize}
Furthermore, the \pt of the system made of the isolated lepton and the \Jpsi
candidate is $\ptlmumu>25$~\GeV and the invariant mass of the system
is restricted to
$10<\mlmumu<160$~\GeV, since the tail of this distribution is more sensitive
to \ttbar and single-top-quark modelling uncertainties.
Finally, a total of \num{12165} candidates remain.%

\begin{figure}[!htb]
\centering
\subfloat[]{\includegraphics[width=0.49\textwidth]{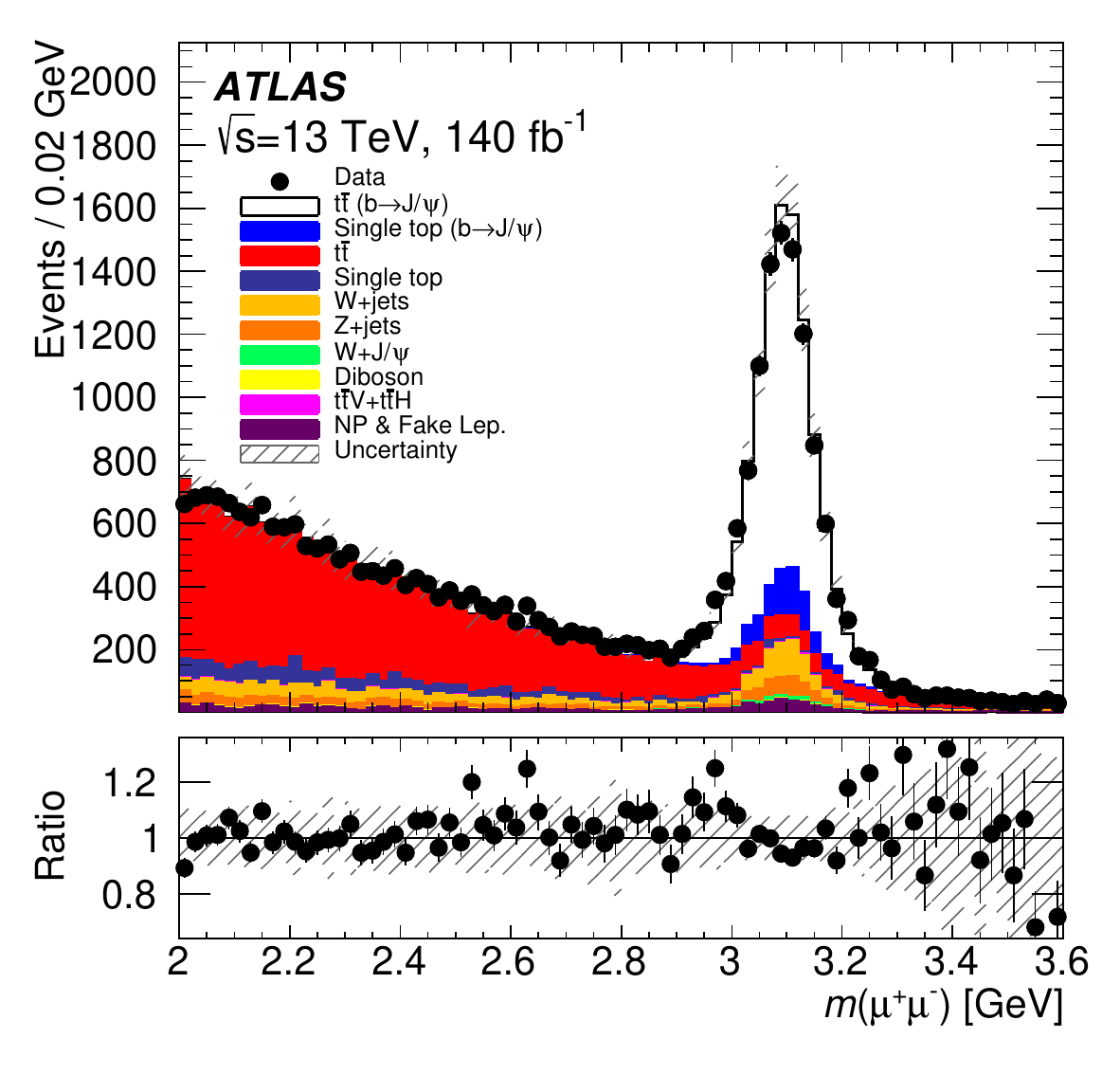}\label{fig:sel:a}}
\subfloat[]{\includegraphics[width=0.49\textwidth]{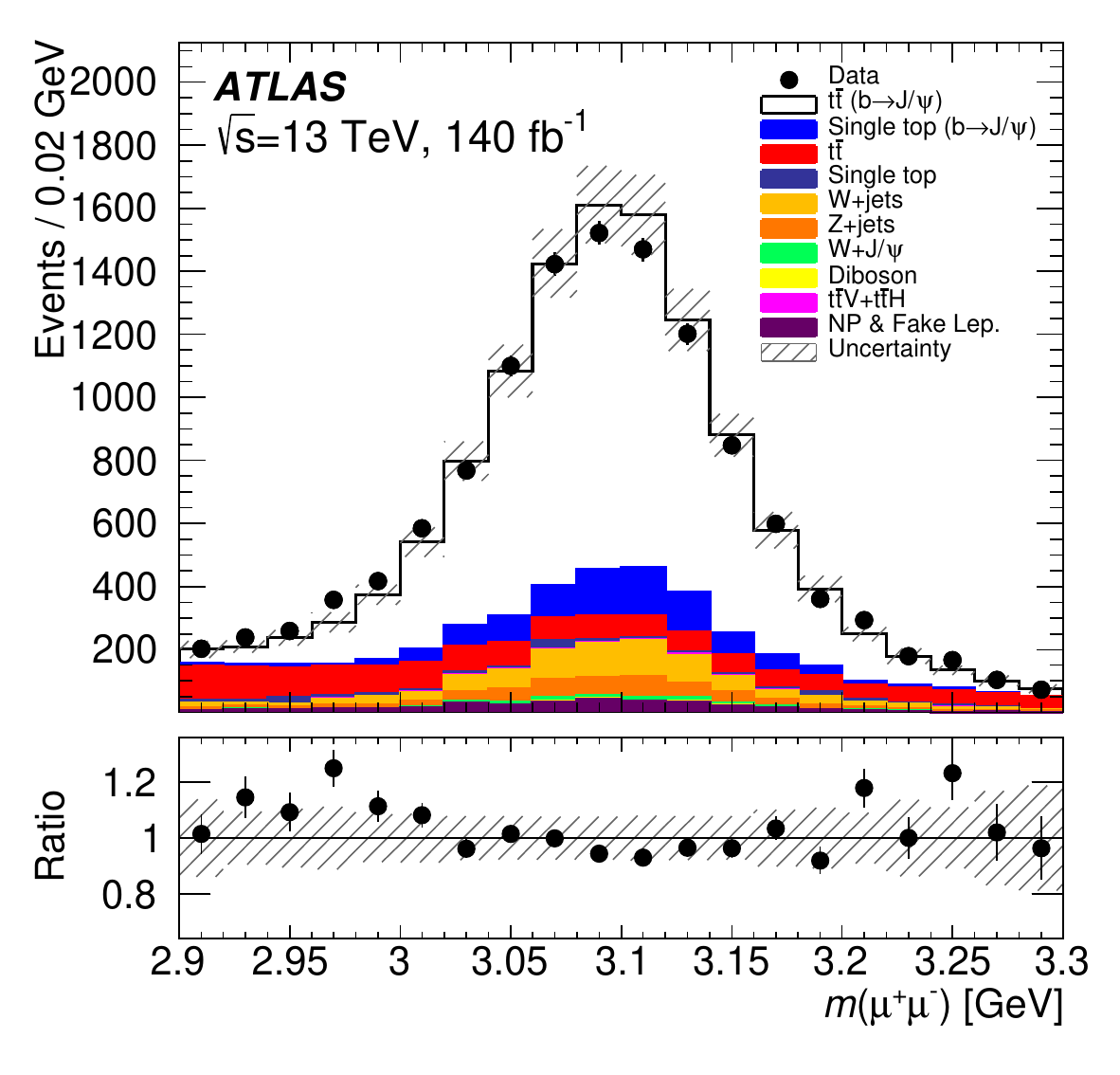}\label{fig:sel:b}}\\
\subfloat[]{\includegraphics[width=0.49\textwidth]{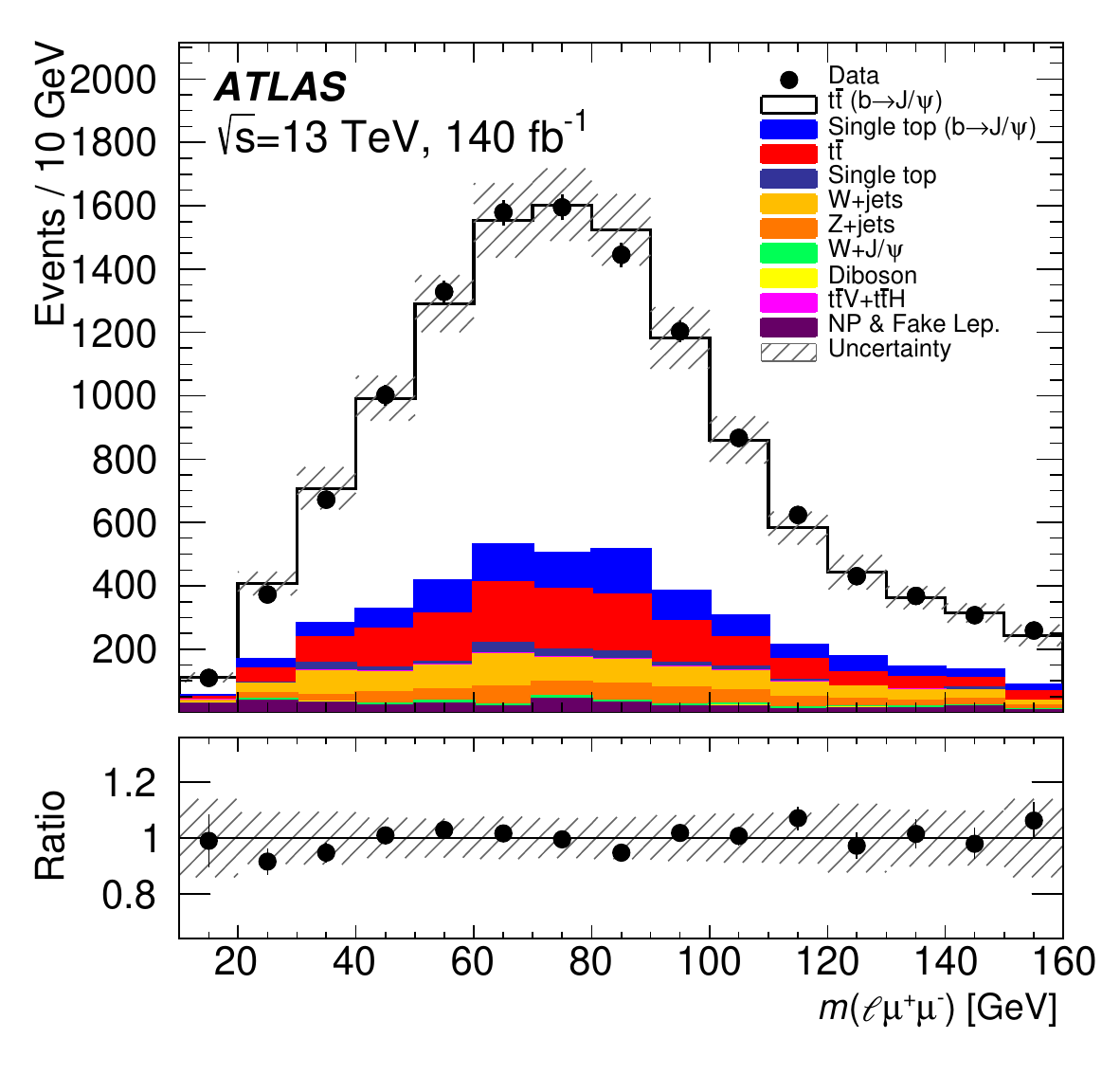}\label{fig:sel:c}}
\subfloat[]{\includegraphics[width=0.49\textwidth]{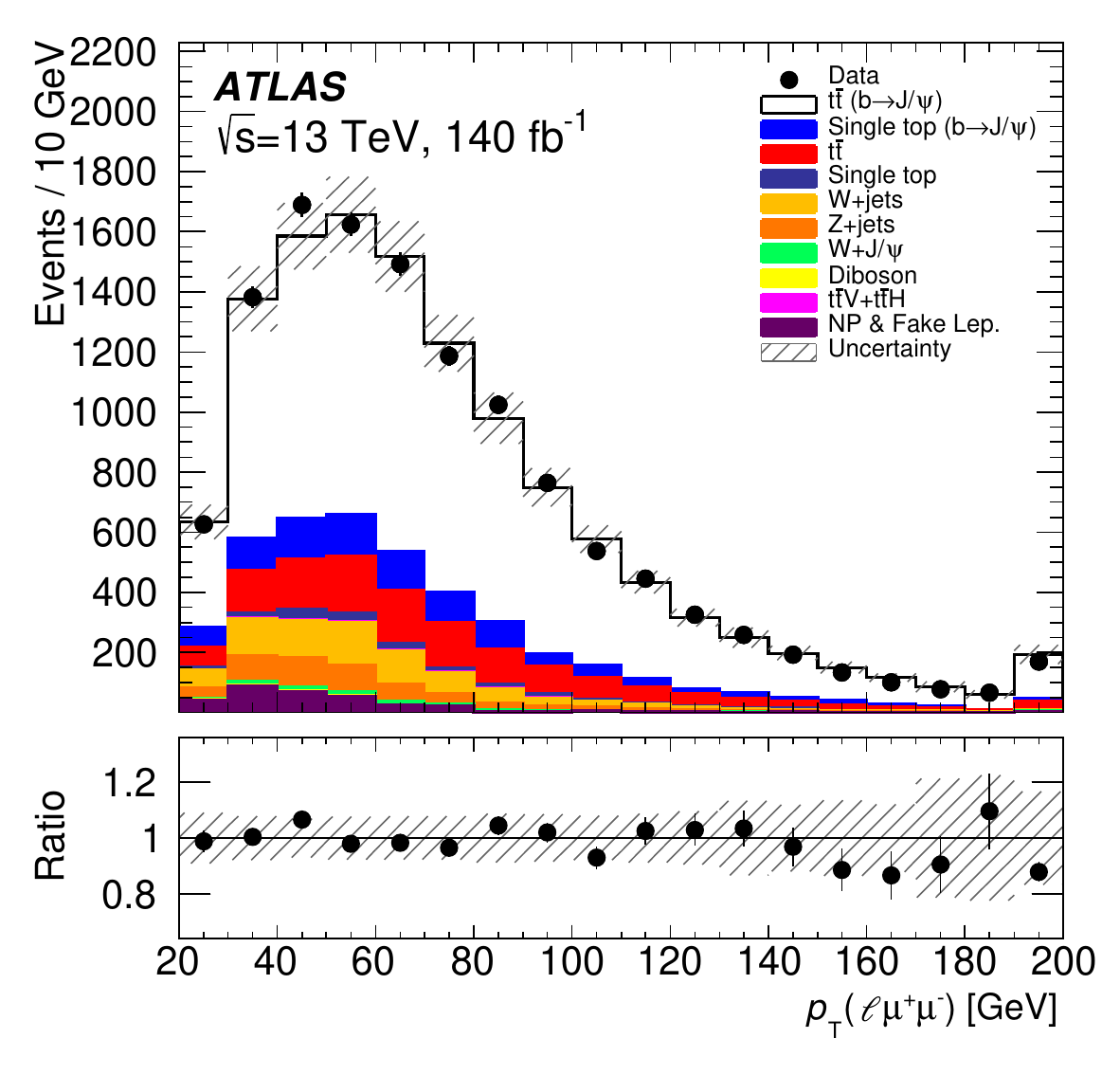}\label{fig:sel:d}}
\caption{Distributions of the invariant mass \mmumu of \Jpsi candidates in
(a) the 2--3.6~\GeV and (b) the 2.9--3.3~\GeV\ mass range,
(c) the invariant mass \mlmumu and
(d) the transverse momentum \ptlmumu of the system made of the selected
isolated lepton and the two muons from the \Jpsi candidate.
The data is shown compared with the expectation from simulation, broken down
into contributions from \ttbar and single-top-quark with and without
$b\rightarrow\Jmm$ decay, \ttVH, \Wjets, \WJpsi, \Zjets,
dibosons and events with non-prompt and fake leptons
(referred to as `NP \& Fake Lep.').
The data is represented as closed circles with statistical uncertainties.
The predictions are shown as solid coloured histograms and are normalised
to the same integrated luminosity as the data.
The shaded area represents the combination of statistical and systematic
uncertainties.
The lower panels show the ratios of the data to the predictions.}
\end{figure}
Figures~\ref{fig:sel:a} and \ref{fig:sel:b} show the distributions of the
invariant mass \mmumu of the \Jpsi candidates for the selected events in an
extended mass window 2--3.6~\GeV and in the final range 2.9--3.3~\GeV,
respectively.
Figures~\ref{fig:sel:c} and \ref{fig:sel:d} present the invariant mass \mlmumu
and the transverse momentum \ptlmumu of the selected isolated lepton $\ell$
and the \Jpsi system.
The prediction reproduces the \mmumu\ distribution in data within the
associated uncertainties, in particular the momentum resolution uncertainty
of the soft-$\mu$, although some differences between the shapes are observed.
These differences have a negligible impact on the analysis.
Table~\ref{tab:yields} shows the number of events in the full \RunTwo data
sample retained after the final selection.
Also shown are the expected numbers of \ttbar and single-top-quark events,
assuming a top-quark mass of $\mtop=172.5$~\GeV, broken down into
contributions with and without the $b\rightarrow\Jmm$ decay, and
other background events, corresponding to the integrated luminosity of the data.
The sum of the predicted number of events agrees within uncertainties with the
observed number of events in the data.
\begin{table}[!hbt]
\centering
\caption{Number of selected events in data after the final selection.
Also shown are the expected numbers of \ttbar and single-top-quark events,
assuming a top-quark mass of $\mtop=172.5$~\GeV, broken down into
contributions with and without the $b\rightarrow\Jmm$ decay, and
other background events, corresponding to the integrated luminosity of the data.
The last two rows show the expected background fraction and the ratio of
observed to expected events.
The total uncertainty includes both statistical and systematic components,
combined in quadrature.}
\label{tab:yields}
\begin{tabular}{    l
S[table-format=5.0]@{\,\( \pm \)\,}
S[table-format=3.0]@{}
}
\toprule
Data                                  & \multicolumn{2}{c}{12\,165} \\
\midrule
\ttbar ($b\rightarrow\Jmm$)           &  7941 &  737 \\
Single-top-quark ($b\rightarrow\Jmm$) &   964 &  109 \\
\midrule
\ttbar                                &  1411 &  145 \\
Single-top-quark                      &   164 &   43 \\
\ttVH                                 &    38 &    8 \\
\Wjets                                &   777 &  258 \\
\Zjets                                &   468 &  138 \\
\WJpsi                                &    78 &   41 \\
diboson                               &    16 &    9 \\
Non-prompt and fake lepton            &   322 &   99 \\
\midrule
Signal+background                     & 12180 &  821 \\
\midrule
\end{tabular}
\begin{tabular}{    l
S[table-format=8.2]@{\,\( \pm \)\,}
S[table-format=1.2]@{}
}
Expected background fraction          & 0.27 &  0.01 \\
Data/(Signal+background)              & 1.00 &  0.07 \\
\bottomrule
\end{tabular}
\end{table}

About 73\% of the selected events come from the signal samples,
containing a \ttbar pair or a single-top-quark with a true
$J/\psi\rightarrow\mu^+\mu^-$ decay from \bquark fragmentation.
Most of these signal events, 90.5\% (7\%), have the two (only one)
soft-$\mu$ coming from the decay of a \Jpsi produced by a $b$- or $c$-hadron
that comes from a \bquark produced by a $t\rightarrow Wb$ decay.
For about 2\% of the signal events, one of the soft-$\mu$ originates from
the prompt \Wboson decay, found near a jet, or radiates a near-collinear
photon mimicking a jet.
For 0.5\% of the signal events, one of the soft-$\mu$ arises from
light-hadron decays or detector background.
These events are referred to as `Fake soft-$\mu$' events in the following.
About 13\% of all selected events contain a \ttbar pair or a single-top-quark
without a true $b\rightarrow J/\psi\rightarrow\mu^+\mu^-$.
The random combination of two selected soft-$\mu$ produces an
exponentially decreasing \mmumu distribution.
Almost all of these combinatorial background events have the two
soft-$\mu$ originating from the decay of a $b$- or $c$-hadron while in
5\% of the cases, one of them is a `Fake soft-$\mu$'.

The contribution from other background processes accounts for
approximately 14\% of the total.
These backgrounds produce a peak in the \mmumu distribution under the
signal peak.
It is dominated by the production of \Wboson and \Zboson boson in association
with jets, for which in about two thirds of the cases a real \Jpsi is produced,
originating from a $b$- or a $c$-jet.
Among \WJpsi selected events, about 70\% originate from double parton
scattering process (DPS) production.
Background events containing non-prompt and fake leptons also lead to
a peak around the \Jpsi.

About half of the \ttbar events have the isolated lepton and
the \Jpsi belonging to the decay of the same top quark,
which is the signature providing the best sensitivity to the top-quark mass.
Putting a requirement on the distance between the isolated lepton
and the \Jpsi could increase the fraction of such good pairings, but
at the cost of a significant loss of events leading to an unacceptable
increase in the statistical uncertainty.


%

\section{Systematic uncertainties}
\label{sec:uncertainties}
The systematic uncertainties that affect the measurements of \mtop
are classified into four categories:
those arising from the number of events in the data and Monte Carlo samples,
from the modelling of \ttbar and single-top-quark processes,
estimated separately, from the modelling of the physics background processes,
and from the effects of the ATLAS detector and reconstruction techniques.
For sources that consist of two-sided variations, a symmetric uncertainty is
estimated as half of the difference between the up and down variations,
assuming that the impact on the measured \mtop is symmetric
around the nominal value.
If both variations point in the same direction compared with the central
result, the maximum is taken as the resulting uncertainty and the
opposite-sign variation is assumed to be of identical size.
In cases where the uncertainty is determined from two MC set-ups,
e.g. for the evaluation of the uncertainty induced by the choice of signal MC
generator, the full observed difference from the nominal result is used
as a symmetric uncertainty.

Since the systematic uncertainties are evaluated based on MC samples of
limited size, each systematic uncertainty has a corresponding statistical
uncertainty, taking into account the statistical
correlation of the samples~\cite{TOPQ-2017-03}.
Using the individual uncertainties
$u_i\pm s_i$ the total uncertainty is calculated as $u\pm s$,
where $u=\sqrt{\sum_i u^{2}_{i}}$
and $s=(1/u)\times\sqrt{\sum_i (u^{2}_{i}\times s^{2}_{i})}$.
The statistical precision from a comparison of two samples $\sigma_{12}$ is
determined for each uncertainty source based on the correlation $\rho_{12}$ of
the underlying samples, using $\sigma_{12}^2=\sigma_{1}^2+\sigma_{2}^2-2\rho_{12}\sigma_1\sigma_2$.
The statistical correlation is expressed as $\rho_{12}=w_{12}/(\sqrt{w_1} \sqrt{w_2})$
with $w_1 = \sum_{i,1} w^{2}_{i}$ and $w_2 = \sum_{i,2} w^{2}_{i}$
the sum of squared weights in the two samples, and
$w_{12} = \sum_{i,12} w^{2}_{i}$ the sum of squared weights
of events present in both samples.
\subsection{Data and Monte Carlo samples}
The limited size of the simulated samples as well as
the non-prompt and fake-lepton background component impacts the
construction of the \mtop-dependent and independent templates.
The corresponding uncertainty is estimated by using a bootstrap
method~\cite{ATL-PHYS-PUB-2021-011}.
The method calibration uncertainty is determined by repeating the mass
extraction on MC samples generated with different values of \mtopgen mass and
estimating the residual bias of the measurement.
A constant is fitted and its statistical uncertainty is assigned as the
uncertainty.
The uncertainty in the integrated luminosity is 0.83\%~\cite{DAPR-2021-01}
and is applied to all processes other than the non-prompt- and
fake-lepton background.
The reweighting of the simulation to match the pile-up distribution in data
involves rescaling the average number of interactions per bunch crossing
to achieve improved agreement between data and simulation for the observed
number of primary vertices.
The uncertainty in determining this rescaling is propagated to the
reweighting factors to estimate the uncertainty due to pile-up.
\subsection{Modelling of \ttbar and single-top-quark processes}
Uncertainties in \ttbar and single-top-quark process modelling include
sources that affect the kinematics of the lepton from the \Wboson boson decay,
as well as the kinematics and fraction of events of different flavour of the
\bhadron giving rise to the \Jmm.
Uncertainties are assessed by either reweighting the nominal samples or by
using the alternative simulation samples described
in Section~\ref{sec:datasimu}.
The full observed difference between the \mtop measurements
is quoted as the systematic uncertainty, unless where mentioned explicitly.

The impact on the measurement from the uncertainties in the \ttbar and
single-top-quark inclusive cross-sections is negligible, as the fit is
performed using only the shape of the distributions and does not depend
on yield variations arising from the assumed top-quark mass.
The uncertainty related to the matching of the NLO calculation is evaluated
using the \ttbar and single-top-quark samples with the alternative \pthard\
setting.
An uncertainty is considered to account for the effect of \ttbar
\NNLO\ corrections, which are known to be not fully covered by the scale
uncertainties of the \ttbar NLO+PS prediction~\cite{CZA-2016,CZA-2017}.
An iterative reweighting procedure is implemented to simulate NNLO+NLL QCD
effects in the parton level $p_{\textrm{T}}^{t}$, $p_{\textrm{T}}^{t\bar{t}}$ and
$m_{t\bar{t}}$ distributions.
The full observed difference using the nominal \ttbar sample with and without
this reweighting is quoted as the systematic uncertainty.
The top-quark decay lineshape uncertainty accounts for the ambiguity in the
models of the on-shell effects in production and compares the two existing
implementations of the smearing and off-shell corrections:
\POWHEG and \MADSPIN.
The uncertainty in the interference between \ttbar and \tW production
at NLO is assessed by comparing the default ‘diagram removal’ scheme with
an alternative ‘diagram subtraction’ scheme~\cite{Frixione:2008yi}.
The full observed difference is quoted as the systematic uncertainty.
Uncertainties in the PS and hadronisation model are assessed by
comparing the \mtop values obtained from the sample generated with
\POWHER[7.1.3] with the nominal \ttbar sample and \POWHER[7.2.1]
for the $t$- and \tW-channels.

Uncertainties in the \bhadron production and decay branching fractions of the
inclusive decays of \bhadrons into \Jmm are derived from the uncertainties in
the rescaling procedure described in Section~\ref{sec:datasimu}
and come entirely from the PDG values.
These uncertainties are propagated through the analysis,
taking into account
the correlation coefficients between production fractions~\cite{Zyla:2020zbs}:
$\textrm{cor(\Bs, \bbaryon)} = 0.064,$
$\textrm{cor(\Bs, \ensuremath{B^{\pm}}=\Bzero)} = -0.633$ and
$\textrm{cor(\bbaryon, \ensuremath{B^{\pm}}=\Bzero)} = -0.813$.
Initial- and final-state QCD radiation (ISR/FSR) lead to a higher jet
multiplicity and different jet energies than the hard process, which affects
the distributions of the observables used in this measurement.
Several systematic variations are made to cover uncertainties in the amount
of ISR radiation.
One uncertainty component is obtained using alternative event weights in
the nominal sample to mimic variations in the \muR and \muF scales in the
matrix elements, varied by factors of \num{0.5} and \num{2.0}, independently.
For another component, additional sets of weights are used to reproduce
the Var3c up/down variants of the A14 tune, altering
the \alphasISR scale~\cite{ATL-PHYS-PUB-2017-007}.
Another uncertainty component is derived from the \ttbar sample with the
alternative value for the \hdamp parameter.
The effect of FSR is evaluated using alternative simulated \ttbar samples
generated with the scale parameters explicitly varied.
The uncertainty from the underlying event is assessed by
using the two samples generated with the Var1 eigentune variation.
Uncertainties in the CR model are assessed by comparing the
two alternative CR model samples (CR1 and CR2) with the CR0 sample.
The largest difference relative to CR0 comes from the CR2 model and is taken
as a symmetric uncertainty.
The uncertainty originating from the PDFs is evaluated using the 30
eigenvectors of the PDF4LHC15 set~\cite{Butterworth:2015oua}.

Uncertainties in the modelling of the \bquark\ fragmentation are evaluated
using samples generated with different \rb values,
by comparing the \mtop values obtained
from the up ($\rb=1.07$) and down ($\rb=1.03$) variations
for the A14--\rb\ tune described in Section~\ref{sec:datasimu}.
To ensure that the estimate based on the \rb values adequately accounts for
the modelling uncertainties in the \bquark\ fragmentation,
two cross-checks are performed using different theoretical
predictions of the fragmentation function.
The first one concerns the variable \xB that is sensitive to the fraction of the
\bquark energy taken by the \bhadron\ in the top-quark rest frame,
when considering the top-quark decay $t\rightarrow Wb$:
\begin{equation*}
\xB = \frac{1}{1- m_{W}^{2}/\mtop^{2} + m_{b}^{2}/\mtop^{2}} \frac{2 p_{B} \cdot p_\textrm{top}}{\mtop^{2}} \, ,
\end{equation*}
where $m_{W}$ and $m_{b}$ are respectively the \Wboson-boson and
the \bquark masses and $p_B$ and $p_\textrm{top}$ are the
\bhadron and the top-quark four momenta~\cite{Corcella:2005,Corcella:2010a}.
Figure~\ref{fig:xbtheory} shows the \xB\ spectrum predicted at
NLO+NLL~\cite{Cacciari:2002,Corcella:2005}, accounting for soft-gluon
resummation in top-quark decays, with the hadronisation according to the
Kartvelishvili model~\cite{Kartvelishvili:1977pi}.
This spectrum is compared with the one obtained at particle level in the \ttbar
\PYTHIA[8] sample, controlled by the parameters of the fragmentation function
and \alphasFSR, which are set in the A14--\rb tune.
The observed differences between the shapes are used to define scale factors
applied on \ttbar events, which translate into variations
of the observed \mlmumu\ distribution.
The second cross-check concerns the prediction of fragmentation functions
at NNLO order accuracy, including NNLL soft gluon resummation
of $B$-hadrons and \Jpsi~\cite{CZA-2023}.
The differences in shape between the differential distributions of
\mljpsi predicted at \NLO\ and \NNLO\ are used to define scale factors
applied to the \ttbar events, which translate into a variation of the
observed \mlmumu\ distribution.
In both cases, the variations in measured \mtop values are comparable in size
to the uncertainty assigned to the \bquark fragmentation modelling.
These variations are not included as dedicated systematic uncertainties
in the final result.
\begin{figure}[htb!]
\begin{center}
\includegraphics[width=0.48\textwidth]{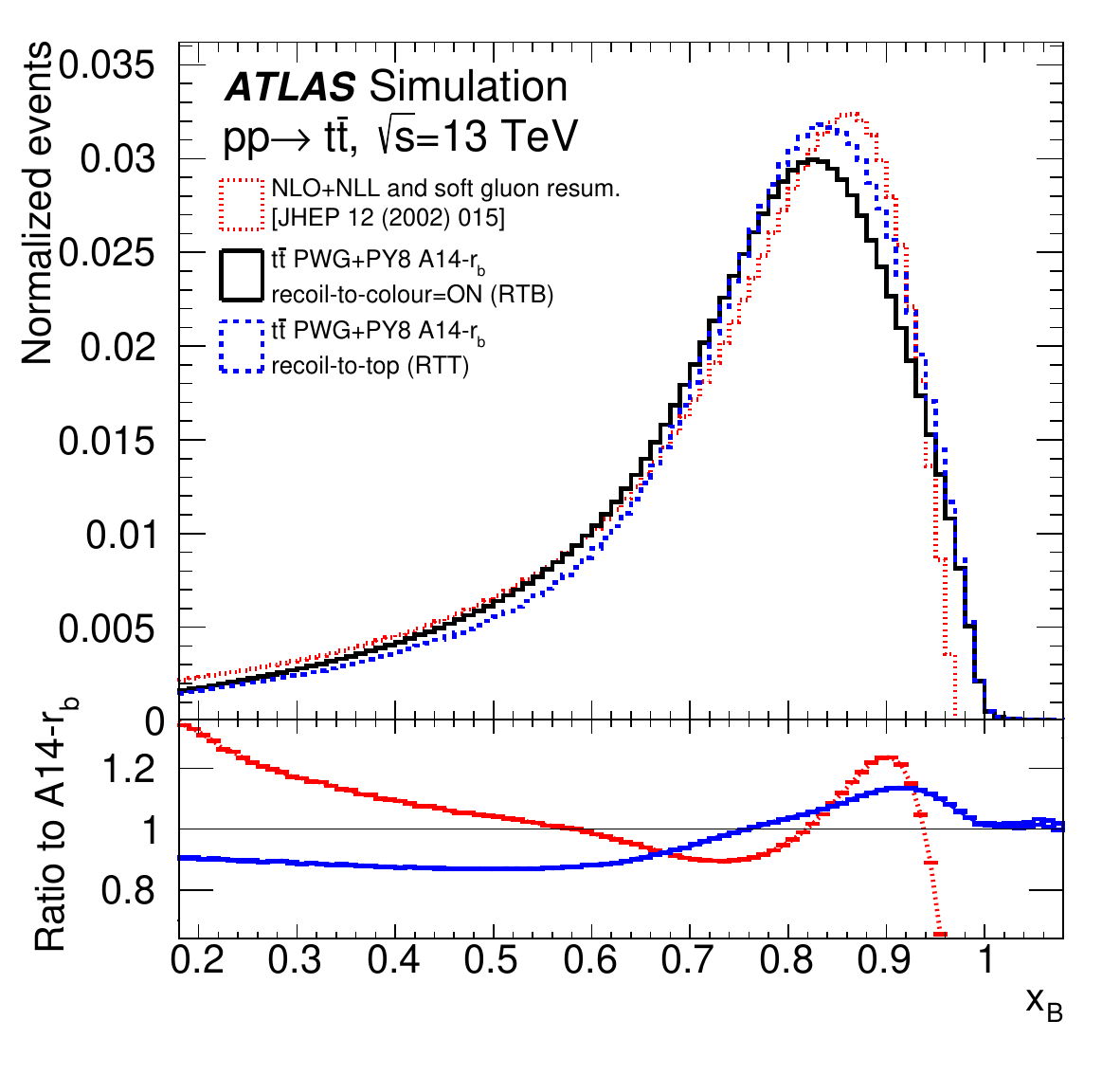}
\end{center}
\caption{Distribution of \xB, the fraction of energy taken by the \bhadron
in the top-quark rest frame, as obtained from theoretical prediction at
NLO+NLL~\cite{Cacciari:2002,Corcella:2005},
accounting for soft-gluon resummation with the hadronisation
according to the Kartvelishvili model~\cite{Kartvelishvili:1977pi},
and obtained at particle level using \ttbar samples.
The \POWPY[8] A14--\rb\ MC samples are based on different recoil schemes using
either the \bquark (`recoil-to-colour=ON') or the top quark
(`recoil-to-top') as the recoil particle.
The lower panel shows the ratios of the theoretical
and `recoil-to-top' predictions to the `recoil-to-colour=ON' one.
\label{fig:xbtheory}
}
\end{figure}

In the simulation of the dipole QCD radiation scheme, there is an ambiguity
in the choice of the recoil particle for the modelling of the second and
subsequent gluon emission from quarks produced by coloured resonance decays,
such as the \bquark in a $t\rightarrow Wb$ process.
Practically this determines how the momentum is re-arranged between the
\Wboson boson and the \bquark, but it has no impact, for example,
on $Z\rightarrow b\bar{b}$ decays~\cite{Brooks:2019xso}.
The default gluon recoil scheme in the nominal \ttbar samples
uses the \bquark as the recoil particle using the \PYTHIA[8] parameter
\textit{recoil-to-colour=ON}, referred to as RTB\@.
The uncertainty due to the ambiguity in the choice of recoil scheme is
assessed by comparing with an alternative \ttbar sample produced
where the top quark is the recoil particle, using the \PYTHIA[8] parameter
\textit{recoil-to-top}, referred to as RTT.
A fully self-consistent variation of the recoil scheme would require a
dedicated tuning of \PYTHIA[8] parameters.
This necessitates measurements in top-quark events with sufficient sensitivity
to this aspect of the modelling that could be fed into existing measurements
able to constrain \alphas and $b$-quark fragmentation.
Despite previous measurements showing sensitivity to the recoil
scheme~\cite{TOPQ-2022-24,TOPQ-2017-17}, and indicating a preference of data
for the recoil scheme used in the nominal simulation of this
analysis~\cite{TOPQ-2022-24}, no such dedicated measurements yet exist.
The full observed difference in \mtop measurements between the RTB and
RTT recoil schemes is $1.07\pm 0.22$~\GeV.
The RTT set-up mildly changes the \Wboson-boson \pt and the
angle between the \Wboson boson and the \bhadron resulting from
the top-quark decay, but it hardens the \bhadron momentum and,
as a consequence, modifies the distribution of \xB,
as shown in Figure~\ref{fig:xbtheory}.
As in Ref~\cite{TOPQ-2017-17},
a reweighting procedure is tested in order to force the \xB spectrum
derived with the RTT scheme to be the same as that for the RTB one.
This would reduce the recoil systematic uncertainty down to 0.10~\GeV\ with a
statistical uncertainty in this uncertainty of 0.21~\GeV.
A fully consistent comparison of the fragmentation description in parton
shower simulations with the available theoretical predictions at
NLO+NLL~\cite{Cacciari:2002,Corcella:2005} and NNLO~\cite{CZA-2023} is beyond
the scope of the present paper.
The full observed difference in \mtop measurements between the RTB and RTT
recoil schemes is therefore conservatively included in the final result,
separately from other systematic modelling uncertainties.
\subsection{Modelling of background processes}
Several sources of uncertainties are considered for the normalisation and shape
of background contributions.
As seen in Section~\ref{sec:sel}, about 13\% of the selected events
originate from \ttbar or single-top-quark events without a true
$b\rightarrow J/\psi\rightarrow\mu^+\mu^-$ leading to a combinatorial background
under the peak of the \Jpsi.
The normalisation and shape of the \mmumu distributions of these contributions
are checked with other selections where this component is enhanced,
in particular by selecting events with three or four jets,
at least one of them being \btagged.
In the \mmumu sideband region [2--2.9]~\GeV where this background is
dominant, the normalisation and shape of the \mlmumu distribution
is also checked.
In all cases, the data is well reproduced by expectations,
and the impact of this effect is found to be negligible.
The fraction of the soft-$\mu$ component of \ttbar events that arises
from `Fake soft-$\mu$', as defined in Section~\ref{sec:sel}, is small,
4\% in the sideband and only 0.5\% in the signal region.
A conservative uncertainty of 25\% in the normalisation of this
component is taken as a systematic uncertainty,
supported by the comparison between data and
prediction in the sidebands of the invariant mass distributions.

Processes like \ttV\ and \ttH\ represent only a small contribution
to the oberved number of events and are included in the background component.
The cross-sections of the \ttV and \ttH processes are known with a precision
around 13\%~\cite{deFlorian:2016spz}.
Still, some processes show discrepancies between data and MC in
previous measurements~\cite{TOPQ-2019-30,TOPQ-2020-20}.
The inclusive cross-sections for diboson production
are calculated using MCFM~\cite{Campbell:2010} and are known
to a precision of better than 10\%~\cite{ATL-PHYS-PUB-2016-002}.
For all these processes, more conservative uncertainties of 50\% are applied
to also cover shape uncertainties and observed differences found in previous
analyses~\cite{ATL-PHYS-PUB-2016-002}.
Uncertainties related to the $W$/\Zjets processes are determined using
two methods.
The uncertainties related to the default \SHERPA samples are determined by
varying independently the \muR and \muF scales by factors of 0.5 and 2 as well
as considering PDF and \alphas variations~\cite{ATL-PHYS-PUB-2017-006}.
The alternative \MGFXFX generator to simulate $W$/$Z$+jet processes,
described in Section~\ref{sec:datasimu}, which provides a different treatment
of multi-jets multiplicity from the nominal samples, leads to
generator-choice-induced differences in the normalisations and shapes of the
distributions \mmumu and \mlmumu, which in turn give rise to
systematic uncertainties.
The uncertainties obtained from the two methods are summed in quadrature
to obtain the full uncertainty for the $W$/$Z$+jets background.
For prompt \WJpsi production, a 50\% normalisation uncertainty is used.
The normalisation and shape of the \mmumu distributions of the peaking
backgrounds are checked in two validation regions, designed to
enrich these components: a region with exactly one reconstructed jet to
validate the \Wjets and \WJpsi production; and one with exactly two jets
and two isolated leptons to validate \Zjets production.
It is verified that the uncertainty arising from differences observed
between data and MC simulation in these validation regions is negligible.

For the data-driven estimate of the non-prompt and fake lepton background,
a 30\% systematic uncertainty in the predicted yields is assigned,
uncorrelated between \ejets and \mjets events.
This uncertainty is supported by comparison of distributions at
low \met values, enriched in non-prompt and fake lepton background events.
\subsection{Modelling of the detector response}
Uncertainties for isolated leptons arise from the efficiency of their
reconstruction, identification and isolation requirements,
all derived from studies using \Zboson\ and
\Jpsi decays~\cite{EGAM-2018-01,MUON-2018-03}.
A similar procedure is used to estimate the trigger
efficiencies~\cite{TRIG-2018-05,TRIG-2018-01}.
The lepton energy/momentum scales and resolutions are obtained from \Zboson and
\Jpsi decays~\cite{EGAM-2021-02,MUON-2022-01}.
The uncertainties associated with the soft-$\mu$ are related in particular to
the muon momentum scale, resolution and identification efficiencies at low \pt.
Although the observable \mlmumu does not involve jets, the various jet
uncertainties impact the analysis through the event selection.
Uncertainties in the jet energy scale (\JES) are evaluated using 29
independent variations in the jet energies that parametrise the uncertainties
in the \JES of $R=0.4$ particle-flow jets~\cite{JETM-2018-05}.
The \JES comprises statistical subcomponents from in situ calibrations,
detector related subcomponents such as energy scales and resolutions of
electromagnetic objects and modelling subcomponents for $\gamma$+jets and
\Zjets calibrations.
The magnitude of the jet energy resolution (\JER) uncertainty is
parameterised in jet \pt and $\eta$~\cite{JETM-2018-05}, and the
uncertainty is propagated by smearing the jet \pt in the simulation.
The uncertainty in the efficiency to satisfy the \JVT requirement is evaluated
by varying the scale factors within their uncertainties~\cite{PERF-2014-03}.
The \btagging efficiencies and mistag rates are measured in data, and scale
factors are derived to correct the predicted tagging rates.
Corresponding uncertainties are taken into account by varying
these scale factors, which depend on jet \pt, $\eta$, and on the hadronic
content of the jet.
They include the uncertainties in the \btagging~\cite{FTAG-2018-01} and
mistagging scale factors~\cite{FTAG-2020-08,FTAG-2019-02}
with nine, four and five eigenvector variations for $b$, $c$/$\tau$-lepton
and light jets, respectively.
The uncertainty in \mtop is derived by varying the scale factors within their
uncertainties and adding the resulting fitted differences in quadrature.
As \MET is not explicitly used in the selection, no uncertainty
is assigned to it.


%

\section{Template method and result}
The analysis method employed in this paper is a template fit.
Templates are simulated distributions of the invariant mass \mlmumu of the
system made of the selected isolated lepton $\ell$ and the two soft-$\mu$
associated with the \Jpsi candidate.
The templates are built from MC samples representing different \mtop values,
with added non-prompt and fake-lepton events.
The templates are split into two components separating the contributions with
a significant \mtop dependence from contributions that do not depend on \mtop.

The \mtop-dependent templates are constructed from \ttbar and
single-top-quark MC samples produced for nine \mtopgen values:
169, 171, 172, 172.25, 172.50, 172.75, 173, 174 and 176~\GeV.
This procedure is adopted, firstly, because \ttbar and single-top-quark
production without $b\rightarrow J/\psi \rightarrow\mu^+\mu^-$ decay,
although formally a background process, still carries information about
the top-quark mass and, secondly, by doing so an \mtop-independent background
template can be used.
A parameterisation of these distributions is performed using the sum of a
Gaussian and a LogNormal function, the latter having the form :
\begin{eqnarray*}
f(x|m_0,k) &=&
\frac{1}{\sqrt{2\pi \cdot \ln(k) \cdot x}}
\exp\left(
\frac{-\ln^2(x/m_0)}{2\ln^2(k)}
\right) \, ,
\end{eqnarray*}
where $m_0$ is the median and $k=\exp(\sigma)$ with $\sigma$ a shape parameter.
The parameterisation is chosen to represent the Gaussian part of the
distribution, coming mostly from \ttbar and single-top-quark events where the
\Jpsi and the prompt lepton come from the same top quark, and a contribution
with a larger tail coming from the cases where the \Jpsi and the isolated lepton
do not come from the same top quark or the \Jpsi candidate does not come from
a \bquark produced by a $t\rightarrow Wb$ decay.
This approach assumes that each fit parameter has a linear
dependence on the top-quark mass, which has been verified.
The \mtop-dependent templates are shown in
Figure~\ref{fig:template:masssensitivity:a}
for three different \mtopgen values: 169, 172.5 and 176~\GeV.
The \mtop-independent template is constructed from the \ttV and \ttH processes
(neglecting their \mtop dependence due to their small contributions),
from the \Wjets, \WJpsi, \Zjets and dibosons processes,
and from events with non-prompt and fake leptons.
A Chebyshev polynomial function~\cite{Abramowitz:1965} is fitted
to this \mtop-independent template as shown in
Figure~\ref{fig:template:masssensitivity:b}.
These functions are shown to provide an adequate parameterisation
for the \mtop-dependent and \mtop-independent templates.
Other choices for the templates fit functions give compatible results.
\begin{figure}[htb!]
\centering
\subfloat[]{\includegraphics[width=0.48\textwidth]{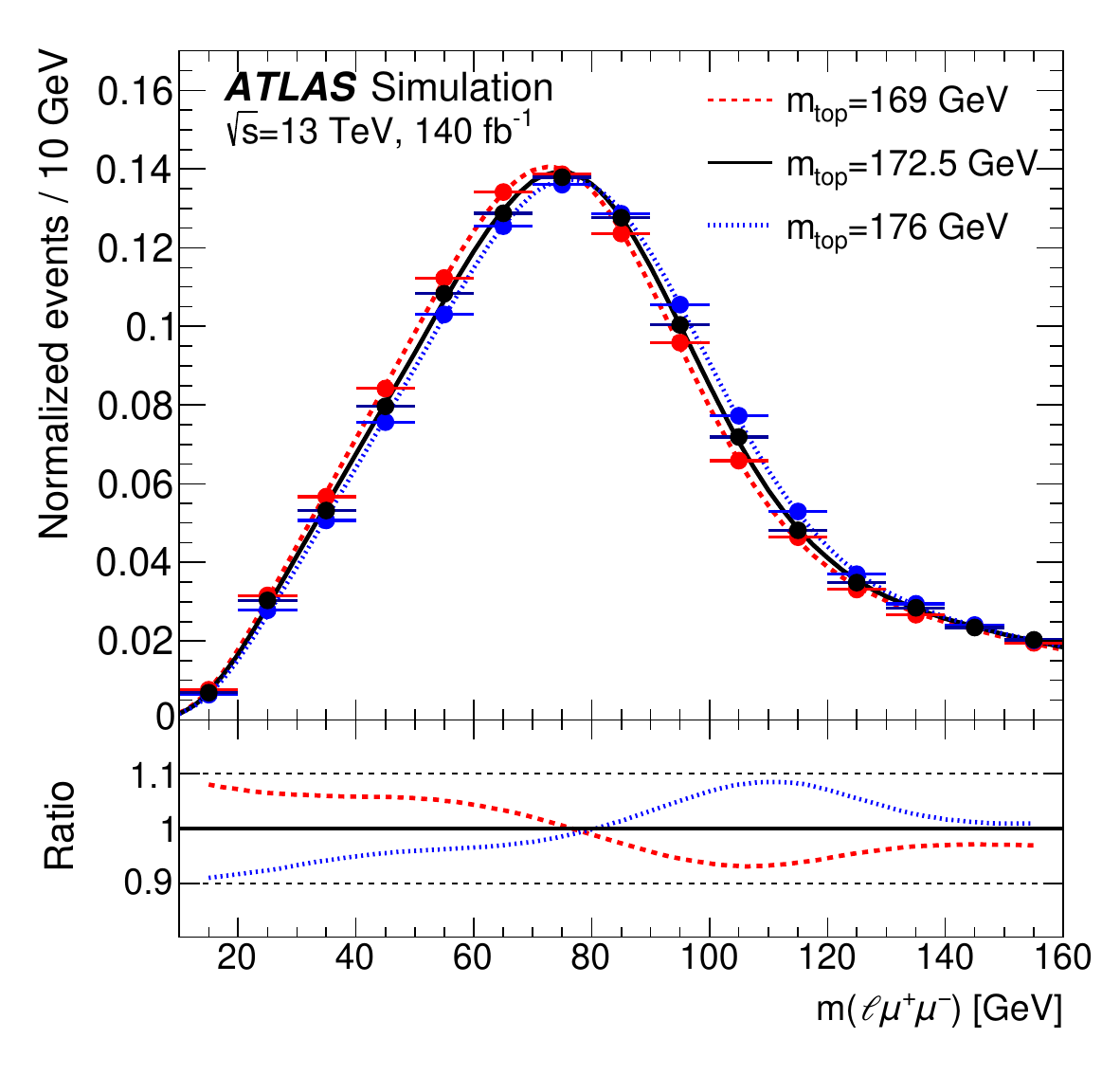}\label{fig:template:masssensitivity:a}}
\subfloat[]{\includegraphics[width=0.48\textwidth]{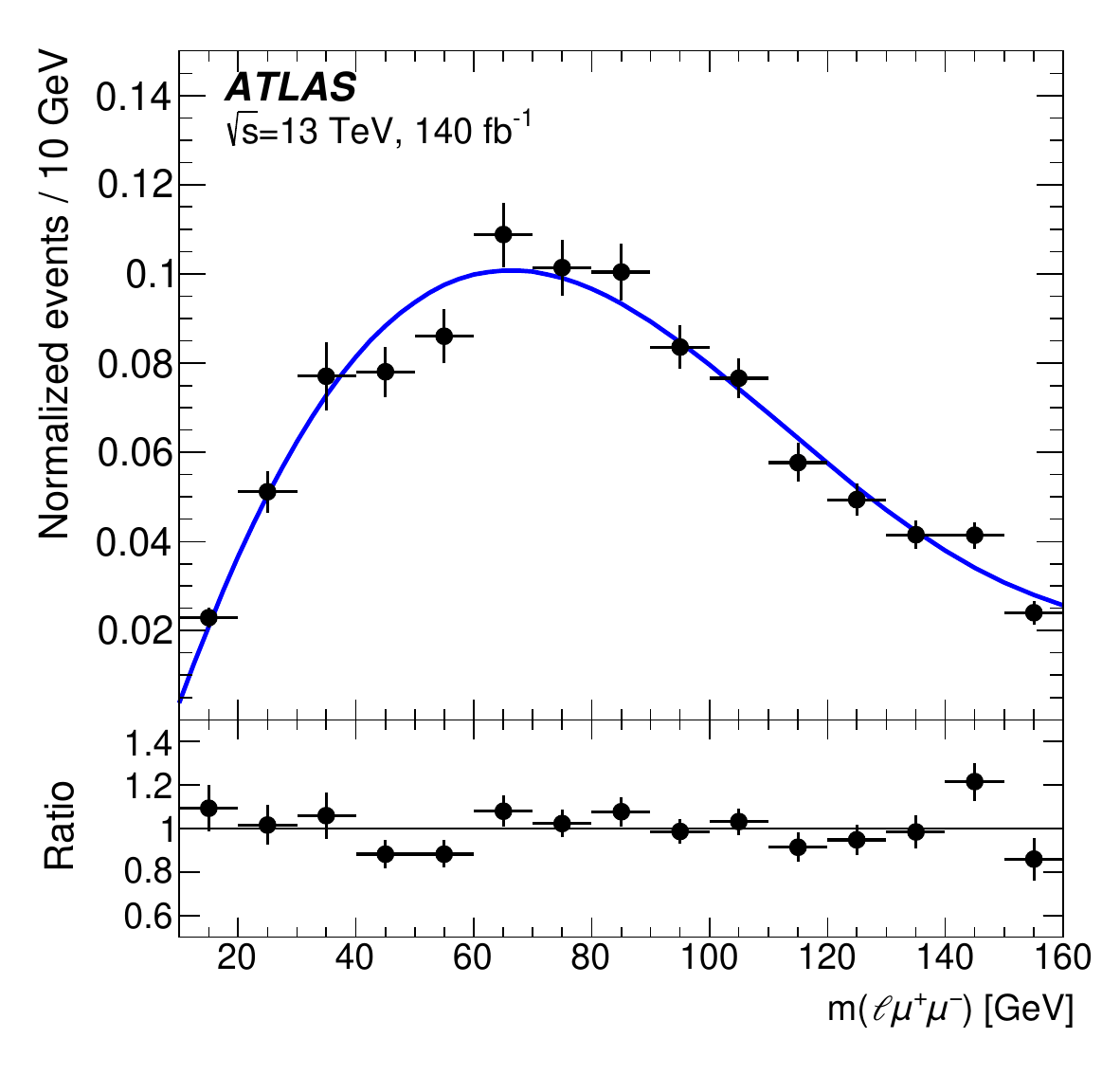}\label{fig:template:masssensitivity:b}}
\caption{Template fit functions for (a) the \mtop-dependent sample shown for
\mtopgen values of 169, 172.5 and 176~\GeV\ and (b) the \mtop-independent
samples compared with the histograms used in the parameterisation.
The lower panel of the \mtop-dependent template plot shows the ratios of the
fit functions obtained with other \mtopgen values to the one obtained
with $\mtopgen=172.5$~\GeV.
The lower panel of the \mtop-independent template plot shows the ratios of the
histogram to the fit function.
}
\end{figure}

The templates are normalised to unity to build \mtop-dependent and
\mtop-independent probability density functions,
respectively $P^\textrm{\mtop-dep.}$ and $P^\textrm{\mtop-indep.}$.
They are summed in an unbinned likelihood that is maximised to give the value
of \mtop that best describes the data:
\begin{equation*}
\mathcal{L}(m_\textrm{top}) = \prod_{i=1}^{N} f_\textrm{\mtop-dep.} P^\textrm{\mtop-dep.}(m^{i}(\ell\mu^+\mu^-)|m_\textrm{top}) + (1 - f_\textrm{\mtop-dep.}) P^\textrm{\mtop-indep.}(m^{i}(\ell\mu^+\mu^-)) \,  ,
\end{equation*}
where $N$ is the number of events and $f_\textrm{\mtop-dep.}$ is the fraction
of events from the \mtop-dependent component.

Figure~\ref{fig:results:data:a} shows the calibration curve,
i.e. the reconstructed \mtop as a function of the generated one \mtopgen.
A linear function is fitted to all mass points and gives
the following parameters:
$\mtop = \left(172.56 \pm 0.10\right) + \left(0.98 \pm 0.06\right) \times (\mtopgen-172.5)$.
The determination of \mtop from these fits is found to be linear and
unbiased relative to the input top-quark mass hypothesis by means
of pseudo experiments~\cite{ATL-PHYS-PUB-2021-011}.
No evidence for a significant bias or non-closure has been found
and the method calibration uncertainty, described in
Section~\ref{sec:uncertainties}, is assigned to cover for any bias below the
sensitivity of the tests performed.


%

The value of \mtop obtained from the fit to data and the corresponding
statistical and systematic uncertainties are:
\begin{equation*}
\mtop = 172.17 \pm 0.80 \, \textrm{(stat)} \pm 0.81 \, \textrm{(syst)} \pm 1.07 \, \textrm{(recoil)}~\GeV \, ,
\end{equation*}
with a total uncertainty of 1.56~\GeV.
The third uncertainty arises from changing the dipole parton shower
gluon-recoil scheme used in top-quark decays.
The data distribution is shown in Figure~\ref{fig:results:data:b} where it
is compared with the predicted distribution and the uncertainty band
corresponding to the total uncertainty reported above.
The uncertainty band is constructed by varying the template fit function
within the statistical and systematic uncertainties.
\begin{figure}[htb!]
\centering
\subfloat[]{\includegraphics[width=0.48\textwidth]{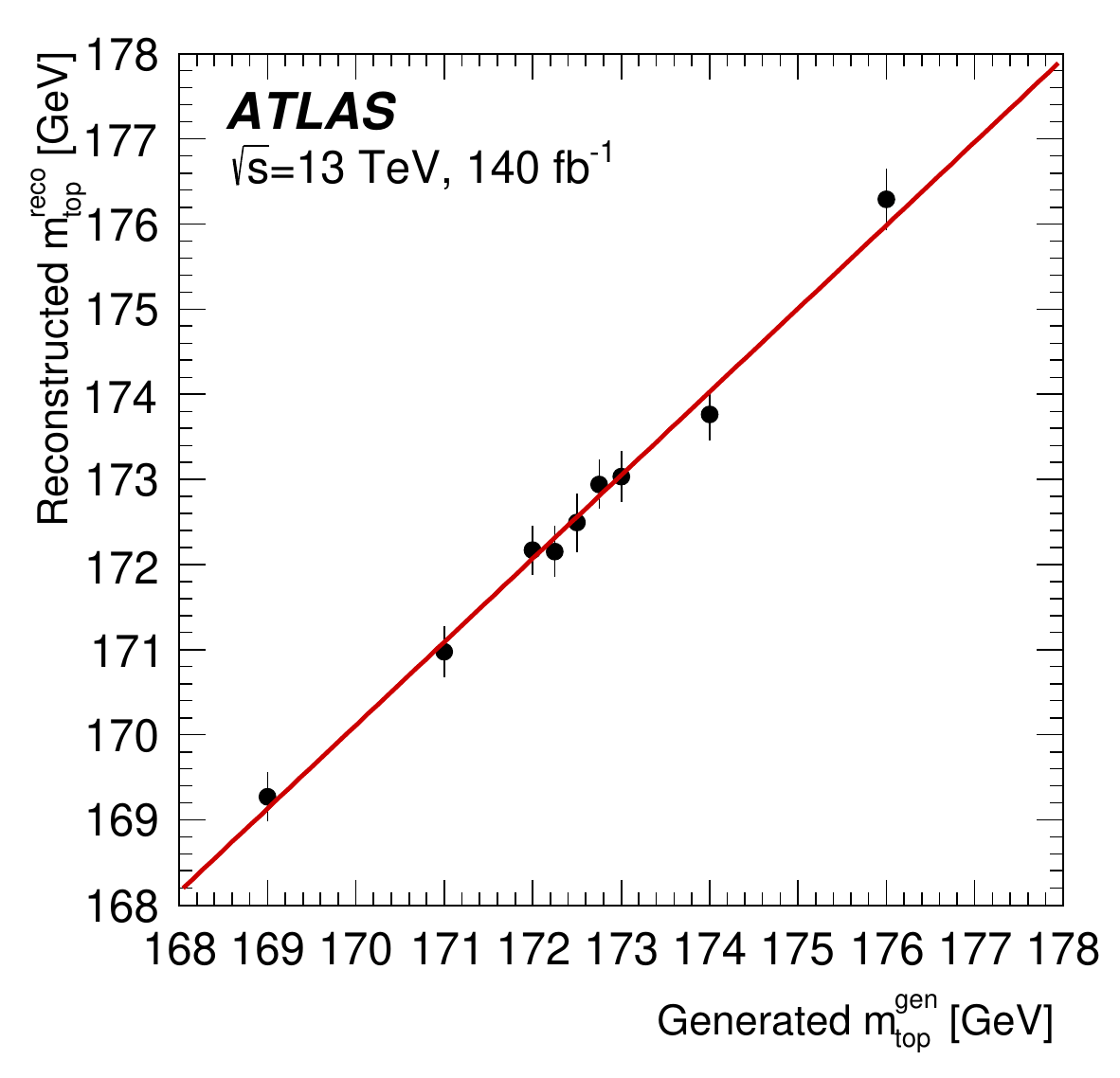}\label{fig:results:data:a}}
\subfloat[]{\includegraphics[width=0.48\textwidth]{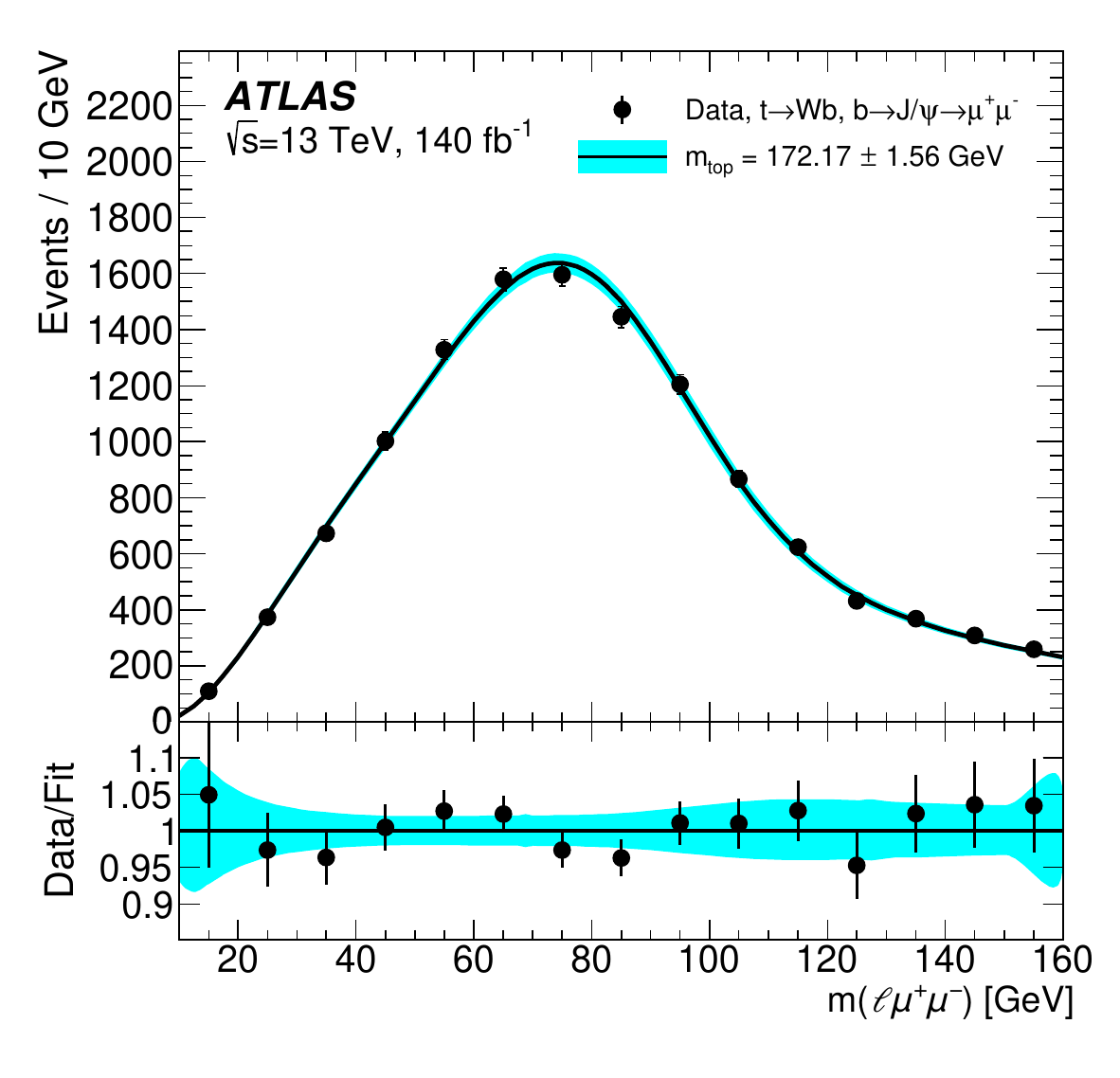}\label{fig:results:data:b}}
\caption{(a) Calibration curve obtained showing the reconstructed \mtop
as a function of the generated one \mtopgen.
The fitted values are represented as closed circles with statistical
uncertainties. A linear function is fitted to the mass points.
(b) The \mlmumu distribution in data compared with the predicted
distribution.
The shaded uncertainty band is constructed by varying the template fit function
within the statistical and systematic uncertainties of the measurement.
The lower panel shows the ratio of data to the template fit function.
}
\end{figure}
The individual uncertainties and their sum in quadrature are
given in Table~\ref{tab:results}.
Uncertainties related to \ttbar and single-top-quark processes are shown
separately and are considered uncorrelated to obtain the total uncertainty.
The statistical and systematic uncertainties, excluding recoil, are at
the same level.
Apart from the dominant recoil uncertainty,
the main sources of systematic uncertainties are due to the parton shower and
hadronisation modelling, final-state radiation modelling, \bquark\ fragmentation
and also large uncertainties from background modelling
in particular from the \Wjets modelling.
The detector uncertainties are on the contrary much smaller; in particular the
uncertainty due to the calibration of the jet energies is below 100~\MeV.
\begin{table}[!hbt]
\centering
\caption{Impact of sources of uncertainty in \mtop.
Each row of the table corresponds to a group of individual
systematic variations.
Uncertainties related to \ttbar and single-top-quark processes are
shown separately and are considered uncorrelated.
For each systematic uncertainty listed, the first value
corresponds to the uncertainty in \mtop,
and the second to the statistical precision of this uncertainty.
The total systematic uncertainty and the corresponding statistical
precision are calculated as discussed
in Section~\ref{sec:uncertainties}.
The total uncertainty is the sum in quadrature of the statistical
and systematic uncertainties.}
\label{tab:results}
\resizebox{0.65\textwidth}{!}{%
\begin{tabular}{lcc}
\toprule
Source of uncertainty                            & Unc. in \mtop[\GeV]  & Stat. precision [\GeV] \\
\midrule
\multicolumn{3}{l}{\textbf{Data and Monte Carlo samples}}                           \\
Statistical error in data                                  & 0.80          &              \\
Statistical error in signal and background model			             & 0.27          & $\pm 0.06$   \\
Method                                           & 0.06          & $\pm 0.10$   \\
Luminosity                                       & 0.01          & $<\pm 0.005$ \\
Pile-up                                          & 0.11          & $\pm 0.02$   \\
\midrule
\multicolumn{3}{l}{\textbf{Modelling of \ttbar process}}                        \\
Matrix element matching                          & 0.14          & $\pm 0.20$   \\
NNLO reweighting                                 & 0.01          & $\pm 0.02$   \\
Top-quark decay lineshape                        & 0.07          & $\pm 0.02$   \\
Parton shower and hadronisation                  & 0.40          & $\pm 0.19$   \\
\bhadron production fractions                    & 0.04          & $\pm 0.02$   \\
\bhadron decay BR                                & 0.08          & $\pm 0.02$   \\
Initial-state QCD radiation                      & 0.10          & $\pm 0.07$   \\
Final-state QCD radiation                        & 0.21          & $\pm 0.05$   \\
Underlying event                                 & 0.20          & $\pm 0.12$   \\
Colour reconnection                              & 0.04          & $\pm 0.28$   \\
Parton distribution function                     & 0.03          & $<\pm 0.005$ \\
\bquark fragmentation \rb                        & 0.15          & $\pm 0.04$   \\
\midrule
\multicolumn{3}{l}{\textbf{Modelling of single-top-quark process}}              \\
Matrix element matching                          & 0.02          & $\pm 0.11$   \\
\ttbar--\tW interference                         & 0.15          & $\pm 0.07$   \\
Parton shower and hadronisation                  & 0.09          & $\pm 0.11$   \\
\bhadron production fractions                    & 0.09          & $\pm 0.07$   \\
\bhadron decay BR                                & 0.02          & $<\pm 0.005$  \\
Initial-state QCD radiation                      & 0.05          & $\pm 0.04$   \\
Final-state QCD radiation                        & 0.06          & $\pm 0.04$   \\
Parton distribution function                     & 0.01          & $<\pm 0.005$ \\
\bquark fragmentation \rb                        & 0.05          & $\pm 0.07$   \\
\midrule
\multicolumn{3}{l}{\textbf{Modelling of background processes}}                  \\
\ttVH, diboson                                   & 0.02          & $\pm 0.01$   \\
\ensuremath{\Wboson/\Zboson\text{+\,jets}}       & 0.40          & $\pm 0.12$   \\
\WJpsi                                           & 0.04          & $<\pm 0.005$ \\
Non-prompt and fake lepton                       & 0.15          & $\pm 0.01$   \\
Fake soft-$\mu$                                  & 0.01          & $<\pm 0.005$ \\
\midrule
\multicolumn{3}{l}{\textbf{Detector response}}                                  \\
Isolated leptons                                 & 0.08          & $\pm 0.01$   \\
Soft-$\mu$                                       & 0.12          & $\pm 0.01$   \\
Light jet energy scale                           & 0.07          & $\pm 0.01$   \\
\bjet energy scale                               & 0.04          & $\pm 0.01$   \\
Jet energy resolution                            & 0.02          & $\pm 0.01$   \\
Flavour tagging                                  & 0.02          & $<\pm 0.005$ \\
\midrule
\midrule
Total systematic uncertainty (excluding recoil)      & 0.81      & $\pm 0.15$      \\
Total stat. and syst. uncertainty (excluding recoil) & 1.14      & $\pm 0.15$      \\
\midrule
Recoil uncertainty                                    & 1.07      & $\pm 0.22$      \\
\midrule
Total uncertainty                                     & 1.56      & $\pm 0.18$      \\
\bottomrule
\end{tabular}
}
\end{table}

A comparison of this result with other \mtop measurements is shown in
Figure~\ref{fig:results:combination}.
This new measurement is found to be in good agreement with previous ATLAS
measurements, the ATLAS+CMS Run 1 combination~\cite{TOPQ-2019-13} and the
CMS measurement using the same final state~\cite{CMS-TOP-15-014}.
Precisely quantifying the level of agreement between the results displayed in
Figure~\ref{fig:results:combination} requires a detailed estimate of the
correlation between each pair of measurements, which is beyond the scope
of this paper.
\begin{figure}[htb!]
\centering
\includegraphics[width=0.8\textwidth]{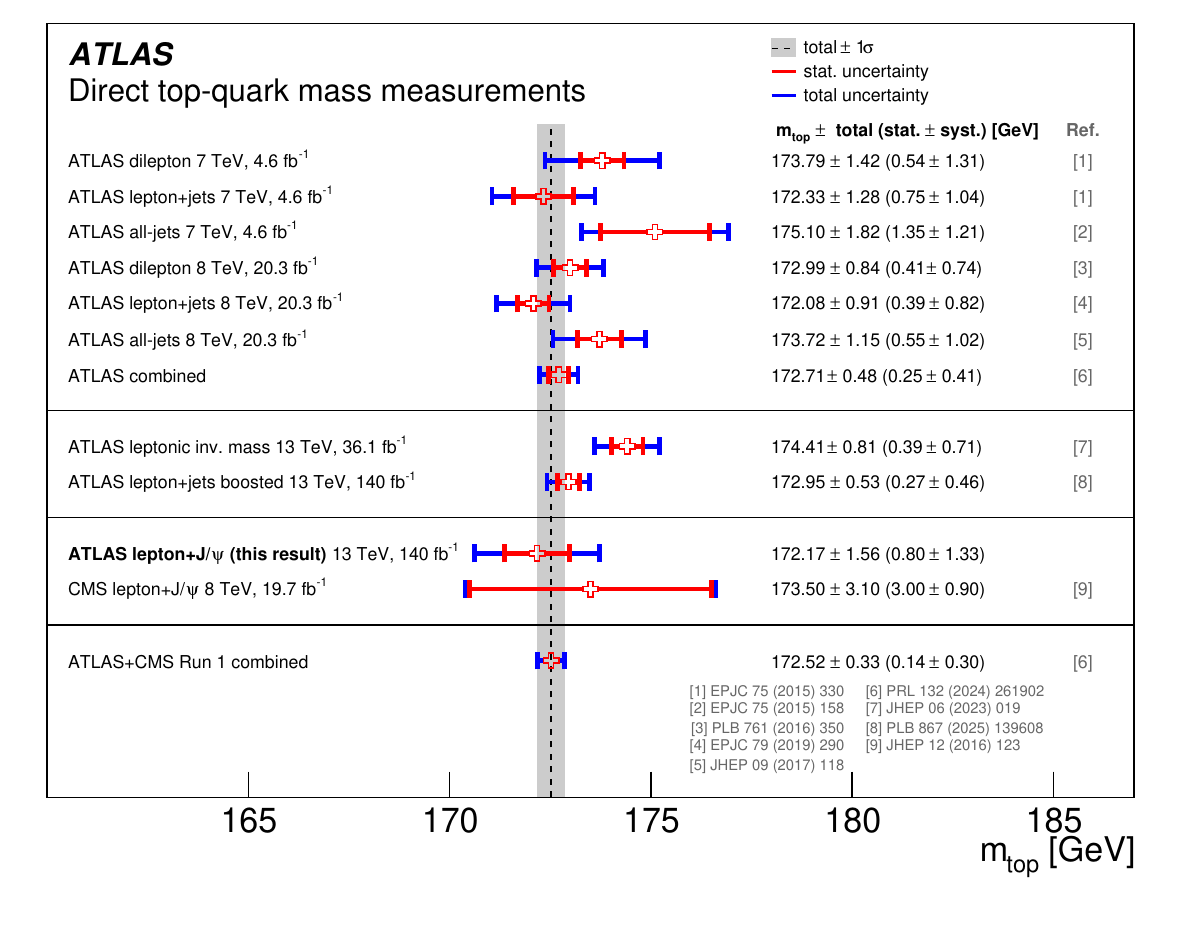}
\caption{This measurement of the top-quark mass \mtop compared with earlier
ATLAS measurements, in particular at 13~\TeV~\cite{TOPQ-2017-17,TOPQ-2022-24},
the ATLAS and CMS Run~1 combination~\cite{TOPQ-2019-13},
and a CMS measurement in the same lepton+\Jpsi channel~\cite{CMS-TOP-15-014}.
The dashed line and shaded band represent the central value and uncertainty of
the ATLAS+CMS Run~1 combination.
}
\label{fig:results:combination}
\end{figure}
%
%
%


%
%
\FloatBarrier

%

\section{Conclusion}
The top-quark mass is measured in top-quark events with a \bquark
decaying into $\Jpsi+X\rightarrow\mu^+\mu^-+X$.
The analysis uses 140~\ifb of proton--proton collision data
recorded during \RunTwo of the LHC at a centre-of-mass of $\rts=13$~\TeV
with the ATLAS detector.
The measurement is based on the invariant mass $m(\ell \mu^+\mu^-)$ of the
system constructed from the isolated lepton $\ell$ from the $W$-boson decay and
the non-isolated $\mu^+\mu^-$ pair from a $J/\psi$ decay of a \bhadron.
An unbinned maximum-likelihood fit to the \mlmumu distribution is
performed to determine the most probable top-quark mass value.
The top-quark mass is measured to be
$\mtop = 172.17 \pm 0.80 \, \textrm{(stat)} \pm 0.81 \, \textrm{(syst)} \pm 1.07 \, \textrm{(recoil)}$~\GeV,
with a total uncertainty of 1.56~\GeV.
The third uncertainty arises from changing the dipole parton shower
gluon-recoil scheme used in top-quark decays.
The main sources of systematic uncertainties are due to the recoil,
parton-shower and hadronisation, \bquark fragmentation, final-state radiation
modelling with uncertainties from background modelling and soft-$\mu$
reconstruction also being significant.
The uncertainty due to the calibration of the jet energies
is smaller and is below 100~\MeV, which is advantageous for future
combinations of this result with those from the other reconstructions
of the top-quark decay products.
The measurement has a large statistical uncertainty, which indicates
that future measurements with larger LHC data samples could reach an improved
precision.
%


%
%
%
%
%
\section*{Acknowledgements}
%

%

%
%

%
%

We thank CERN for the very successful operation of the LHC and its injectors, as well as the support staff at
CERN and at our institutions worldwide without whom ATLAS could not be operated efficiently.

The crucial computing support from all WLCG partners is acknowledged gratefully, in particular from CERN, the ATLAS Tier-1 facilities at TRIUMF/SFU (Canada), NDGF (Denmark, Norway, Sweden), CC-IN2P3 (France), KIT/GridKA (Germany), INFN-CNAF (Italy), NL-T1 (Netherlands), PIC (Spain), RAL (UK) and BNL (USA), the Tier-2 facilities worldwide and large non-WLCG resource providers. Major contributors of computing resources are listed in Ref.~\cite{ATL-SOFT-PUB-2026-001}.

We gratefully acknowledge the support of ANPCyT, Argentina; YerPhI, Armenia; ARC, Australia; BMWFW and FWF, Austria; ANAS, Azerbaijan; CNPq and FAPESP, Brazil; NSERC, NRC and CFI, Canada; CERN; ANID, Chile; CAS, MOST and NSFC, China; Minciencias, Colombia; MEYS CR, Czech Republic; DNRF and DNSRC, Denmark; IN2P3-CNRS and CEA-DRF/IRFU, France; SRNSFG, Georgia; BMFTR, HGF and MPG, Germany; GSRI, Greece; RGC and Hong Kong SAR, China; ICHEP and Academy of Sciences and Humanities, Israel; INFN, Italy; MEXT and JSPS, Japan; CNRST, Morocco; NWO, Netherlands; RCN, Norway; MNiSW, Poland; FCT, Portugal; MNE/IFA, Romania; MSTDI, Serbia; MSSR, Slovakia; ARIS and MVZI, Slovenia; DSI/NRF, South Africa; MICIU/AEI, Spain; SRC and Wallenberg Foundation, Sweden; SERI, SNSF and Cantons of Bern and Geneva, Switzerland; NSTC, Taipei; TENMAK, T\"urkiye; STFC/UKRI, United Kingdom; DOE and NSF, United States of America.

Individual groups and members have received support from BCKDF, CANARIE, CRC and DRAC, Canada; CERN-CZ, FORTE and PRIMUS, Czech Republic; COST, ERC, ERDF, Horizon 2020, ICSC-NextGenerationEU and Marie Sk{\l}odowska-Curie Actions, European Union; Investissements d'Avenir Labex, Investissements d'Avenir Idex and ANR, France; DFG and AvH Foundation, Germany; Herakleitos, Thales and Aristeia programmes co-financed by EU-ESF and the Greek NSRF, Greece; BSF-NSF and MINERVA, Israel; NCN and NAWA, Poland; La Caixa Banking Foundation, CERCA Programme Generalitat de Catalunya and PROMETEO and GenT Programmes Generalitat Valenciana, Spain; G\"{o}ran Gustafssons Stiftelse, Sweden; The Royal Society and Leverhulme Trust, United Kingdom; United States of America.

In addition, individual members wish to acknowledge support from CERN: European Organization for Nuclear Research (CERN DOCT); Chile: Agencia Nacional de Investigaci\'on y Desarrollo (FONDECYT 1230812, FONDECYT 1240864, Fondecyt 3240661, Fondecyt Regular 1240721); China: Chinese Ministry of Science and Technology (MOST-2023YFA1605700, MOST-2023YFA1609300), National Natural Science Foundation of China (NSFC - 12175119, NSFC 12275265); Czech Republic: Czech Science Foundation (GACR - 24-11373S), Ministry of Education Youth and Sports (ERC-CZ-LL2327, FORTE CZ.02.01.01/00/22\_008/0004632), PRIMUS Research Programme (PRIMUS/21/SCI/017); EU: H2020 European Research Council (ERC - 101002463); European Union: European Research Council (BARD No. 101116429, ERC - 948254, ERC 101089007), European Regional Development Fund (HE COFUND GA No.101081355, ERDF), European Union, Future Artificial Intelligence Research (FAIR-NextGenerationEU PE00000013), Italian Center for High Performance Computing, Big Data and Quantum Computing (ICSC, NextGenerationEU); France: Agence Nationale de la Recherche (ANR-21-CE31-0013, ANR-21-CE31-0022, ANR-22-EDIR-0002, ANR-24-CE31-0504-01); Germany: Deutsche Forschungsgemeinschaft (DFG - 469666862, DFG - CR 312/5-2); China: Research Grants Council (GRF); Italy: Istituto Nazionale di Fisica Nucleare (ICSC, NextGenerationEU), Ministero dell'Università e della Ricerca (NextGenEU 153D23001490006 M4C2.1.1, NextGenEU I53D23000820006 M4C2.1.1, NextGenEU I53D23001490006 M4C2.1.1, SOE2024\_0000023); Japan: Japan Society for the Promotion of Science (JSPS KAKENHI JP22H01227, JSPS KAKENHI JP22H04944, JSPS KAKENHI JP22KK0227, JSPS KAKENHI JP24K23939, JSPS KAKENHI JP24KK0251, JSPS KAKENHI JP25H00650, JSPS KAKENHI JP25H01291, JSPS KAKENHI JP25K01023); Norway: Research Council of Norway (RCN-314472); Poland: Ministry of Science and Higher Education (IDUB AGH, POB8, D4 no 9722), Polish National Science Centre (NCN 2021/42/E/ST2/00350, NCN OPUS 2023/51/B/ST2/02507, NCN OPUS nr 2022/47/B/ST2/03059, NCN UMO-2019/34/E/ST2/00393, UMO-2022/47/O/ST2/00148, UMO-2023/49/B/ST2/04085, UMO-2023/51/B/ST2/00920, UMO-2024/53/N/ST2/00869); Portugal: Foundation for Science and Technology (FCT); Spain: Agencia de Gestión de Ayudas Universitarias y de Investigación (AGAUR - 2023 BP 00141), Generalitat Valenciana (ASFAE/2022/008), Ministry of Science and Innovation (MCIN \& NextGenEU PCI2022-135018-2, MICIN \& FEDER PID2021-125273NB, RYC2019-028510-I, RYC2020-030254-I, RYC2021-031273-I, RYC2022-038164-I), Ministerio de Ciencia, Innovación y Universidades/Agencia Estatal de Investigaci\'on (PID2022-142604OB-C22); Sweden: Carl Trygger Foundation (Carl Trygger Foundation CTS 22:2312), Swedish Research Council (Swedish Research Council 2023-04654, VR 2021-03651, VR 2022-03845, VR 2022-04683, VR 2023-03403, VR 2024-05451), Knut and Alice Wallenberg Foundation (KAW 2018.0458, KAW 2022.0358, KAW 2023.0366); Switzerland: Swiss National Science Foundation (SNSF - PCEFP2\_194658); United Kingdom: Royal Society (NIF-R1-231091); United States of America: U.S. Department of Energy (ECA DE-AC02-76SF00515), Neubauer Family Foundation.

%
%


%
\clearpage
\printbibliography
\clearpage
 
\begin{flushleft}
\hypersetup{urlcolor=black}
{\Large The ATLAS Collaboration}

\bigskip

\AtlasOrcid[0000-0002-6665-4934]{G.~Aad}$^\textrm{\scriptsize 104}$,
\AtlasOrcid[0000-0001-7616-1554]{E.~Aakvaag}$^\textrm{\scriptsize 17}$,
\AtlasOrcid[0000-0002-5888-2734]{B.~Abbott}$^\textrm{\scriptsize 123}$,
\AtlasOrcid[0000-0002-0287-5869]{S.~Abdelhameed}$^\textrm{\scriptsize 119a}$,
\AtlasOrcid[0000-0002-1002-1652]{K.~Abeling}$^\textrm{\scriptsize 55}$,
\AtlasOrcid[0000-0001-5763-2760]{N.J.~Abicht}$^\textrm{\scriptsize 49}$,
\AtlasOrcid[0000-0002-8496-9294]{S.H.~Abidi}$^\textrm{\scriptsize 30}$,
\AtlasOrcid[0009-0003-6578-220X]{M.~Aboelela}$^\textrm{\scriptsize 45}$,
\AtlasOrcid[0000-0002-9987-2292]{A.~Aboulhorma}$^\textrm{\scriptsize 36e}$,
\AtlasOrcid[0000-0001-5329-6640]{H.~Abramowicz}$^\textrm{\scriptsize 157}$
% ...
\bigskip
\\

% ...
$^{17}$Department for Physics and Technology, University of Bergen, Bergen; Norway.\\
% ...
$^{30}$Physics Department, Brookhaven National Laboratory, Upton NY; United States of America.\\
% ...
$^{36}$$^{(a)}$Facult\'e des Sciences Ain Chock, Universit\'e Hassan II de Casablanca;$^{(b)}$Facult\'{e} des Sciences, Universit\'{e} Ibn-Tofail, K\'{e}nitra;$^{(c)}$Facult\'e des Sciences Semlalia, Universit\'e Cadi Ayyad, LPHEA-Marrakech;$^{(d)}$LPMR, Facult\'e des Sciences, Universit\'e Mohamed Premier, Oujda;$^{(e)}$Facult\'e des sciences, Universit\'e Mohammed V, Rabat;$^{(f)}$Institute of Applied Physics, Mohammed VI Polytechnic University, Ben Guerir; Morocco.\\
% ...
$^{45}$Physics Department, Southern Methodist University, Dallas TX; United States of America.\\
% ...
$^{49}$Fakult\"{a}t Physik, Technische Universit{\"a}t Dortmund, Dortmund; Germany.\\
% ...
$^{55}$II. Physikalisches Institut, Georg-August-Universit\"{a}t G\"ottingen, G\"ottingen; Germany.\\
% ...
$^{104}$CPPM, Aix-Marseille Universit\'e, CNRS/IN2P3, Marseille; France.\\
% ...
$^{119}$$^{(a)}$New York University Abu Dhabi, Abu Dhabi;$^{(b)}$United Arab Emirates University, Al Ain; United Arab Emirates.\\
% ...
$^{123}$Homer L. Dodge Department of Physics and Astronomy, University of Oklahoma, Norman OK; United States of America.\\
% ...
$^{157}$Raymond and Beverly Sackler School of Physics and Astronomy, Tel Aviv University, Tel Aviv; Israel.\\
% ...
$^{a}$ Also at Affiliated with an institute formerly covered by a cooperation agreement with CERN.\\
$^{b}$ Also at An-Najah National University, Nablus; Palestine.\\
$^{c}$ Also at Borough of Manhattan Community College, City University of New York, New York NY; United States of America.\\
$^{d}$ Also at Center for Interdisciplinary Research and Innovation (CIRI-AUTH), Thessaloniki; Greece.\\
$^{e}$ Also at Centre of Physics of the Universities of Minho and Porto (CF-UM-UP); Portugal.\\
$^{f}$ Also at CERN, Geneva; Switzerland.\\
% ...

\end{flushleft}


%

%
%

%

%
%
%
%

\end{document}